\documentclass[a4paper,fleqn,usenatbib]{mnras}
\usepackage[T1]{fontenc}
\usepackage{ae,aecompl}
\usepackage{graphicx}	
\usepackage{amsmath}	
\usepackage{amssymb}	
\usepackage{booktabs}	
\usepackage[normalem]{ulem}
\usepackage{pdflscape}
\usepackage{url}
\usepackage{verbatim}

\usepackage{newtxtext,newtxmath}

\usepackage{tablefootnote,footnotehyper} 
\usepackage{array}

\newcolumntype{H}{>{\setbox0=\hbox\bgroup}c<{\egroup}@{}} 

\title[The $\delta$\,Sct stars of Cep--Her]{The $\delta$\,Scuti stars of the Cep--Her Complex. I: Pulsator fraction, rotation, asteroseismic large spacings, and the $\nu_{\rm max}$ relation}

\author[Simon J. Murphy et al.]{Simon J. Murphy,$^{1}$\thanks{E-mail: simon.murphy@usq.edu.au (SJM)} Timothy R. Bedding,$^{2}$ Anuj Gautam,$^{1}$ \and Ronan P. Kerr,$^{3}$ and Prasad Mani$^{2}$
\\
$^{1}$ Centre for Astrophysics, University of Southern Queensland, Toowoomba, QLD 4350, Australia\\
$^{2}$ Sydney Institute for Astronomy, School of Physics, University of Sydney, Sydney NSW 2006, Australia\\
$^{3}$ Department of Astronomy, University of Texas at Austin, 2515 Speedway, Stop C1400, Austin, Texas, USA 78712-1205
}

\date{Accepted XXX. Received YYY; in original form ZZZ}
\pubyear{2024}
\begin{document}
\label{firstpage}
\pagerange{\pageref{firstpage}--\pageref{lastpage}}
\maketitle

\begin{abstract}
We identify delta Scuti pulsators amongst members of the recently-discovered Cep--Her Complex using light curves from the Transiting Exoplanet Survey Satellite (TESS). We use Gaia colours and magnitudes to isolate a subsample of provisional Cep--Her members that are located in a narrow band on the colour--magnitude diagram compatible with the zero-age main sequence. 
The $\delta$\,Sct pulsator fraction amongst these stars peaks at 100\% and we describe a trend of higher pulsator fractions for younger stellar associations. 
We use four methods to measure the frequency of maximum amplitude or power, $\nu_{\rm max}$, to minimise methodological bias and we demonstrate their sound performance. The $\nu_{\rm max}$ measurements display a correlation with effective temperature, but with scatter that is too large for the relation to be useful. We find two ridges in the $\nu_{\rm max}$--$T_{\rm eff}$ diagram, one of which appears to be the result of rapid rotation causing stars to pulsate in low-order modes. We measure the $\nu_{\rm max}$ values of $\delta$\,Sct stars in four other clusters or associations of similar age (Trumpler~10, the Pleiades, NGC\,2516, and Praesepe) and find similar behaviour with $T_{\rm eff}$.
Using \'echelle diagrams we measure the asteroseismic large spacing, $\Delta\nu$, for 70 stars, and find a correlation between $\Delta\nu$, rotation, and luminosity that allows rapid rotators seen at low inclinations to be distinguished from slow rotators. We find that rapid rotators are more likely than slow rotators to pulsate, but they do so with less regular pulsation patterns.
We also investigate the reliability of Gaia's {\tt vbroad} measurement for A-type stars, finding that it is mostly accurate but underestimates $v\sin i$ for slow rotators ($v\sin i < 50$\,km\,s$^{-1}$) by 10--15\%.
\end{abstract}




\section{Introduction}
\label{sec:intro}

Asteroseismology of star clusters has long held great promise. The idea is simple: with ensemble asteroseismology, one aims to use multiple stars to achieve the same modelling goal, most commonly a cluster age measurement \citep{basuetal2011}. Historically, the motivation arose from it being much easier to measure distances to clusters than to lone stars \citep{breger2000}. Now, with Gaia \citep{gaiacollaboration2023a}, we can measure distances to individual stars accurately \citep{bailer-jonesetal2021},\footnote{Incidentally, those distances can be verified asteroseismically for red giants, e.g. \citet{zinnetal2019, khanetal2023}.} and also use those distances and kinematics to establish cluster membership \citep{pangetal2022,hunt&reffert2023,kerretal2023}. In addition, clusters can now be treated with hierarchical Bayesian modelling, where not only are the stars assumed to have the same target distribution in age and metallicity, but each individual star also contributes to the overall prior \citep{olivaresetal2018,lyttleetal2021}.

Cluster asteroseismology has delivered several significant results in recent years, especially from the study of red giants (RGs). \citet{brogaardetal2023} found empirical evidence for the long-standing expectation that rotational mixing and core overshooting affect main-sequence lifetimes, and hence cluster age measurements, via the study of RGs in NGC\,6866 (see also \citealt{hidalgoetal2018}). It is widely known that ancient globular clusters have subpopulations or have had multiple generations of star formation (see the review by \citealt{bastian&lardo2018}), and asteroseismic evidence of this has been found in the globular clusters M4 \citep{tailoetal2022} and M80 \citep{howelletal2024}, as well as in some massive open clusters \citep{sandquistetal2020}. Clusters have also been used to verify the asteroseismic scaling relations for RGs \citep[e.g.][]{brogaardetal2016}, while asteroseismology has been able to identify former (escaped) members of open clusters (\citealt{brogaardetal2021}; see also \citealt{heyletal2022,beddingetal2023}).

The pulsations of intermediate-mass main-sequence members of star clusters have been somewhat less studied, except perhaps for the broad attention historically given to blue stragglers \citep{bailyn1995,pychetal2001,templetonetal2002,jeonetal2004,porettietal2008,mcnamara2011,stepienetal2017}. With the Transiting Exoplanet Survey Satellite (TESS) collecting photometry of most of the sky, asteroseismology of A stars on or near the main-sequence ($\delta$\,Sct stars) in clusters has undergone a renaissance, most notably for the Pleiades, but also for Praesepe, $\alpha$\,Per, and others \citep{kerretal2022a,kerretal2022b,murphyetal2022,beddingetal2023,pamosortegaetal2022,pamosortegaetal2023}. This is no surprise, given that clusters might hold the answers to important questions that are especially pertinent to A-type stars, such as the origin and efficiency of angular momentum transport in stars \citep{aertsetal2019,denhartoghetal2020}, the origin and prevalence of chemical peculiarities with age \citep{palla&stahler2000,stepien2000,gray&corbally2002,fossatietal2007,smalleyetal2017}, and how the $\delta$\,Sct pulsator fraction changes with age \citep{murphyetal2019,beddingetal2023}. The latter is a topic we address in this work.

In this work we also study the $\nu_{\rm max}$--$T_{\rm eff}$ relation for $\delta$\,Sct stars.
It has long been known that hotter $\delta$\,Sct stars tend to pulsate in higher radial overtones \citep{breger&bregman1975}. While a $\nu_{\rm max}$ relation is well-tested and widely exploited in solar-like oscillators \citep[e.g.][]{stelloetal2009b,coelhoetal2015,yuetal2018}, such a relation for $\delta$\,Sct stars has so far produced correlations with large scatter \citep{bowman&kurtz2018,barcelofortezaetal2018,barcelofortezaetal2020,hasanzadehetal2021}. Those studies focussed on analysing large datasets of \textit{Kepler} and TESS photometry for heterogeneous samples of stars, so it remains unclear whether the scatter is a consequence of the broad range of encompassed physical parameters (most notably metallicity and age), or from idiosyncrasies of the driving and damping of pulsations in $\delta$\,Sct stars, which might pose a fundamental limit on the available precision of the relation. A recent change in direction has been to study a homogeneous sample, such as that offered by 36 $\delta$\,Sct stars in the Pleiades star cluster \citep{beddingetal2023}, for which a $\nu_{\rm max}$ scaling relation seems doubtful, and 35 $\delta$\,Sct stars in NGC\,2516 \citep{glietal2024}, wherein a $\nu_{\rm max}$ relation appears for about a dozen stars within a narrow colour range. In this work, we use 195 $\delta$\,Sct stars in the \mbox{Cep--Her} association to isolate a homogeneous sample in metallicity and age with which to revisit the question of whether a $\nu_{\rm max}$--$T_{\rm eff}$ relation exists for $\delta$\,Sct stars.

\mbox{Cep--Her} contains at least four kinematically distinct associations or subgroups \citep{kerretal2024}, and is sometimes referred to as the \mbox{Cep--Her} Complex. We use the latter term only when the scope of different subgroups is relevant. All subgroups within the Complex are younger than $\sim$100\,Myr (\citeauthor{kerretal2024}; and Sec.\,\ref{ssec:members} below). There are no red giants or other viable solar-like oscillators with which to determine asteroseismic ages. The stars are also younger than the empirical limit of gyrochronology \citep{boumaetal2023,luetal2024}. Existing age solutions in Cep-Her are computed almost entirely using isochronal ages, which can vary by more than a factor of two depending on the choice of model (\citealt{herczeg&hillenbrand2015}; \citealt{kerretal2024}). As such, new methods with more reliable absolute scaling are necessary to better constrain the ages of the many substructures in Cep-Her. Independent asteroseismic ages for these subgroups would be valuable, and will be the topic of Paper II. In this work, we identify the Cep--Her members that pulsate as $\delta$\,Sct stars, collate their physical parameters, and measure their asteroseismic large spacing $\Delta\nu$ where possible.


\section{Methodology}
\label{sec:methods}

\subsection{Stellar parameters and association membership}
\label{ssec:members}

Our stellar sample is based on the \citet{kerretal2023} Cep--Her membership list. This catalogue includes basic stellar properties for all candidate members including Gaia photometry, distances from \citet{bailer-jonesetal2021}, reddening estimates derived from the \citet{lallementetal2019} reddening maps, and membership probability estimates $P_{\rm mem}$ based on the relative populations of nearby photometrically young and old stars. We required $P_{\rm mem}>0.25$, removing probable non-members from the sample. The photometry, distances, and reddening can be combined to produce absolute values of Gaia colour and absolute magnitude. We required de-reddened Gaia colour $(G_{\rm BP} - G_{\rm RP})_0 < 0.6$, as well as Gaia absolute magnitude $M_G < 5$, limiting the population to the range of temperatures where pulsators are typically found, while using the cut on $M_G$ to remove dim stars with poor photometry, as well as any white dwarfs. From a list of 986 Gaia targets after these cuts, we successfully cross-matched TESS Input Catalogue (TIC) numbers for 958 stars, constituting our final sample. The other 28 stars had been resolved into duplicated TIC numbers.

The Cep--Her Complex has recently been found to contain four sub-components with distinct dynamics and ages: Cinyras, Orpheus, Cupavo, and Roslund~6 \citep{kerretal2024}. The Cinyras and Orpheus Associations have PARSEC isochronal ages that span 28-43\,Myr and 25-40\,Myr, respectively, while  while Cupavo has ages between 54 and 80 Myr and Roslund 6 may be older than 100 Myr. Of the stars in this sample, 205 lie in Cinyras, 478 are in Orpheus, 249 are in Cupavo, and 54 are in Roslund\,6. The sample therefore spans only a small range of ages corresponding to $<$5\% of the main-sequence lifetime. 

We used the Gaia Data Release 3 (DR3) IDs to extract the Renormalised Unit Weight Error ({\tt ruwe}) from the Gaia DR3 catalogue \citep{gaiacollaboration2023a}. High {\tt ruwe} values are indicative of unresolved binaries \citep{evans2018,rizzutoetal2018,belokurovetal2020,stassun&torres2021,penoyreetal2022}, but not with 100\% reliability (\citealt{gallenneetal2023}; Dholakia et al. in prep.). We also note that circumstellar disk material can inflate {\tt ruwe} values for single stars \citep{fittonetal2022}, so {\tt ruwe} values for stars in young associations might be inflated on average when compared to older populations, even after accounting for the dynamical evaporation of binary systems on timescales of hundreds of Myr \citep{fujiietal2012}. We therefore stopped short of assigning a binarity flag based on {\tt ruwe} values, but we did consider {\tt ruwe} values as approximate indicators of binarity in conjunction with other observations where appropriate.

Also from Gaia DR3, we extracted the {\tt vbroad} and {\tt vbroad\_error} values where available. Since A stars often rotate rapidly, with a mean $v\sin i$ in excess of 100\,km\,s$^{-1}$  \citep{royeretal2007}, the total line broadening is dominated by rotation. Hence, rapid rotators in our sample will have large values of {\tt vbroad} (unless seen from low inclinations) and slow rotators will not (see also \citealt{gootkinetal2024}). Rotation causes a deformation that scales as the square of the angular rotation frequency, ($\Omega_{\rm rot}^2$). The resulting decrease in stellar density has two consequences important to this work: firstly, it means that rotating models are required to model these $\delta$\,Sct stars (see discussion in \citealt{murphyetal2022}); and secondly, the corresponding decrease in effective temperature pushes rapid rotators to the right (and sometimes up, depending on inclination) on the colour-magnitude diagram \citep{perezhernandezetal1999, beddingetal2023}. We evaluate the reliability of {\tt vbroad} for $\delta$\,Sct stars in Sec.\,\ref{ssec:vbroad}.

Stellar effective temperatures ($T_{\rm eff}$) and their uncertainties were taken from the TESS Input Catalogue (TIC; \citealt{stassunetal2019}). The TIC temperatures were calculated from spectroscopy originating from nine catalogues covering over 3 million sources, and were supplemented with photometry where required (see \citealt{stassunetal2019} for details). For our targets, the median uncertainty is 156\,K, which is rather low compared to the calibration of the $T_{\rm eff}$ scale with interferometry \citep{casagrandeetal2014,whiteetal2018}. However, these uncertainties are merely indicative and do not underpin our analysis. Note that we did not use Gaia DR3 {\tt gspphot} temperatures because 19\:per\:cent of our sample had no available values in this catalogue. The situation was worse for {\tt gspspec} temperatures ($>$90\:per\:cent missing). Instead, we used both TIC $T_{\rm eff}$ and de-reddened Gaia colour, and we provide them with other stellar parameters online and in the Appendix (Table\:\ref{tab:big}) for the subset of stars analysed in detail in this work (sample selection is described in Sec.\,\ref{ssec:variables} and Sec.\,\ref{ssec:pack}). The Gaia colour is a reliable and widely-available temperature proxy.


\subsection{Downloading TESS lightcurves}
\label{ssec:lks}

We used light curves from the Transiting Exoplanet Survey Satellite (TESS; \citealt{rickeretal2015}). These lightcurves are now available in a variety of cadences. During the first two years of the TESS mission, stars were observed either in 2-min cadence, or were available through custom lightcurves created from Full Frame Images (FFIs) taken at 30-min cadence. In the first extended mission, the FFI cadence reduced to 10-min \citep{bell2020}, and these light curves are now available for many stars as official data products made by the Science Processing Operations Centre (SPOC; \citealt{jenkinsetal2016}). The FFI cadence was reduced again to 200\,s in the second extended mission. A 20-s cadence is also available for a smaller number of stars. 

In this work, we used 2-min cadence light curves when available, and the 10-min FFI light curves otherwise, using the SPOC light curves in both cases. We did not mix cadences for any individual target, did not create any custom lightcuves, and used neither the 20-s nor the 200-s cadences as there were comparatively few Cep--Her targets available in these cadences. Where multiple sectors were available for a given star at the same cadence, we combined those sectors without any adjustment of the times \citep{bedding&kjeldsen2022}. Most of the light curves we collated have data from one sector (24\%) or two sectors (69\%), with a minority having three (1\%) or four (5\%) sectors. All lightcurves were retrieved with the {\sc lightkurve} package \citep{lightkurvecollaboration2018}.


\subsection{Identifying variable stars}
\label{ssec:variables}

To identify $\delta$\,Sct variables in the association, we calculated the amplitude spectrum for each light curve with the Lomb--Scargle periodogram \citep{lomb1976,scargle1982} implemented in {\sc astropy} \citep{astropy2022}. For stars with only 10-min FFIs we used the Nyquist frequency (72\,d$^{-1}$) as the upper frequency limit of the calculation, whereas for the 2-min lightcurves we used 90\,d$^{-1}$. In both cases the lower limit was 10\,d$^{-1}$ and the frequency resolution was set as $1/(4T)$, where $T$ is the timespan of the data. The lower limit was justified via both the Period--Luminosity relation and stellar models, which indicate the fundamental radial mode should lie above about 15\,d$^{-1}$ \citep{ziaalietal2019,baracetal2022,murphyetal2023}.

\begin{figure}
\begin{center}
\includegraphics[width=0.48\textwidth]{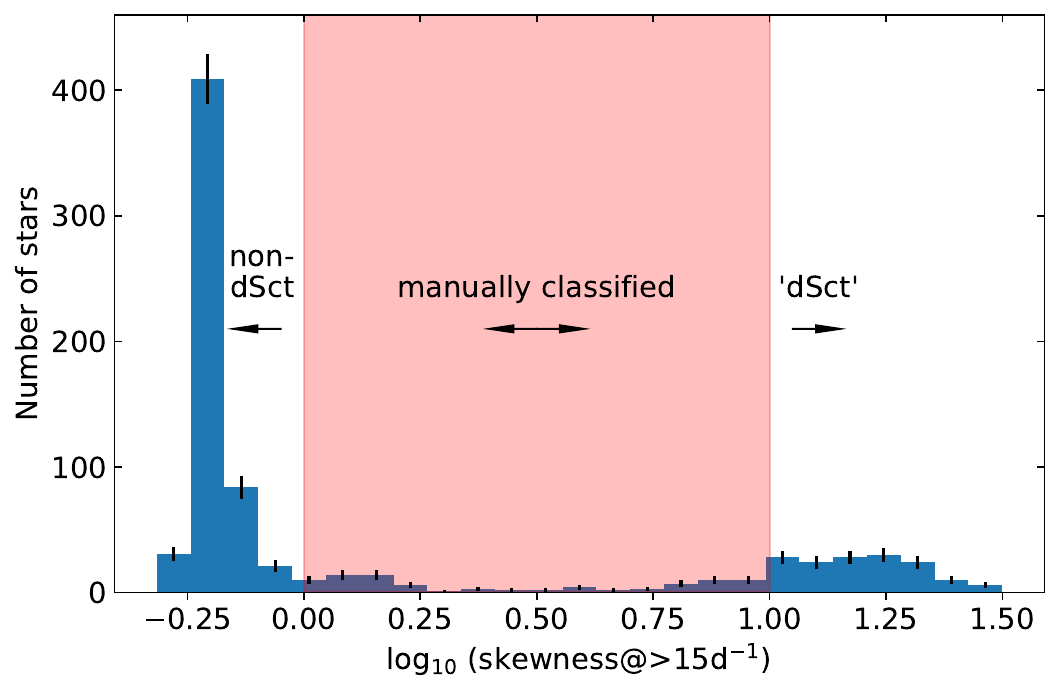}\\
\caption{Distribution of the skewness of the amplitude spectra of the stars in  \mbox{Cep--Her}. The sample was automatically classified based on skewness, and manually classified for stars with log (skew) between 0.0 and 1.0. $\sqrt{N}$ error bars are shown.}
\label{fig:skewness}
\end{center}
\end{figure}

Stars whose amplitude spectra have high skewness are most likely pulsators \citep{murphyetal2019,barbaraetal2022, readetal2024}, whereas low skewness is suggestive of white noise. We measured the skewness of the amplitude array at frequencies above 15\,d$^{-1}$. Given the potential for spectral leakage or harmonics, this lower limit in frequency was necessary to distinguish $\delta$\,Sct stars from other classes whose variability is found at lower frequencies, such as SPB stars, $\gamma$\,Dor stars, and eclipsing binaries. We found the distribution of the resulting skewness values was bimodal, comprising non-variables with $\log_{10} ({\rm skewness}) \lesssim 0.0$, and $\delta$\,Sct stars with $\log_{10} ({\rm skewness}) \gtrsim 1.0$ (Fig.\,\ref{fig:skewness}). We automatically assigned a `dSct' flag to all 188 stars with $\log_{10} ({\rm skewness}) > 0.5$, but visually checked the 88 stars with $0.0 < \log_{10} ({\rm skewness}) < 1.0$ and reassigned any misclassifications (16 additional stars given `dSct' flag, 9 stars had their flag removed). Ambiguous cases were left with their original (automated) assignations. The result was 195 stars labelled as `dSct'.

\begin{figure}
\begin{center}
\includegraphics[width=0.48\textwidth]{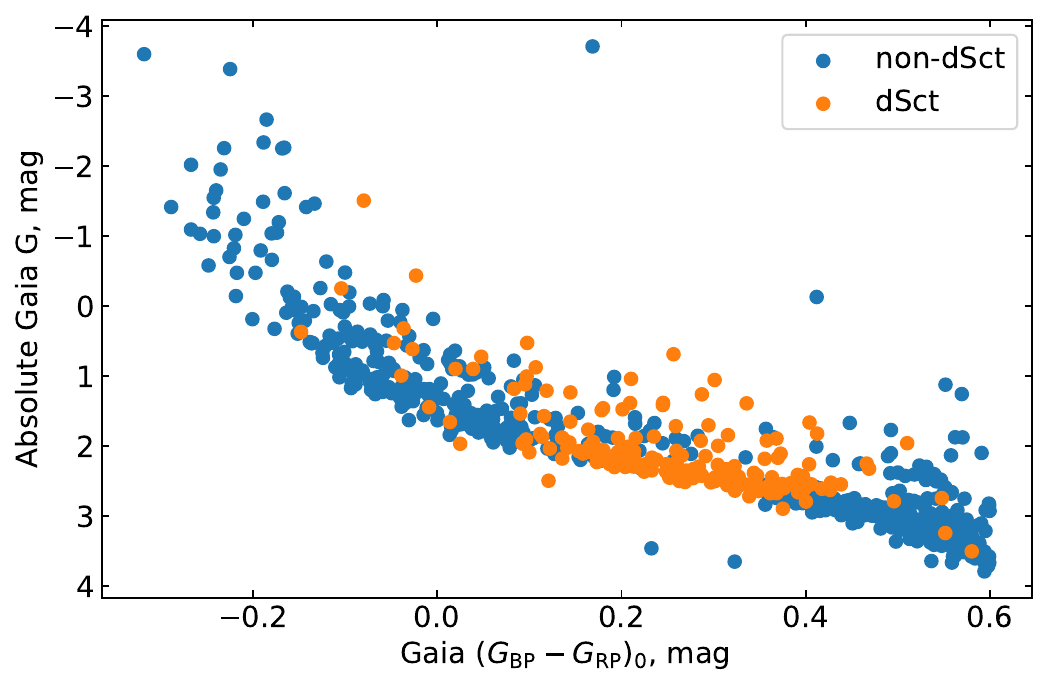}\\
\includegraphics[width=0.48\textwidth]{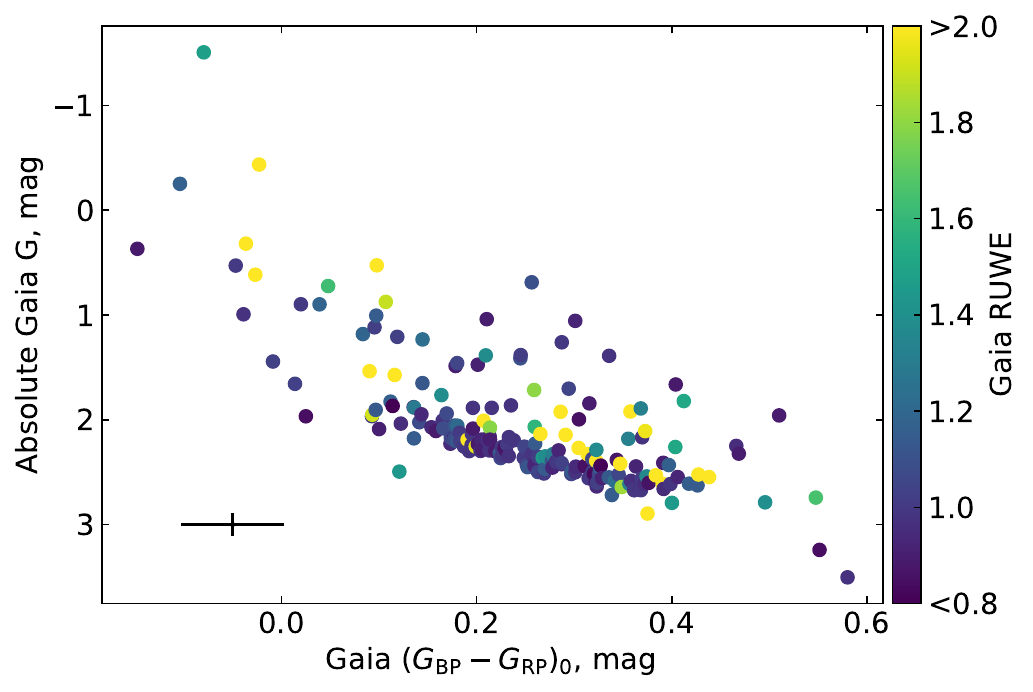}
\caption{Top: Colour--magnitude diagram of the \mbox{Cep--Her} association, also showing all stars in the sample with the `dSct' flag. Absolute magnitude is given in Gaia $G$ band, and colour is the de-reddened $(G_{\rm BP} - G_{\rm RP})_0$ colour. Bottom: The same colour--magnitude diagram, for the `dSct' stars only, with plot symbols colour-coded by the Gaia {\tt ruwe} parameter.}
\label{fig:CMD}
\end{center}
\end{figure}

Fig.\,\ref{fig:CMD} shows the colour--magnitude diagram of  \mbox{Cep--Her} with the $\delta$\,Sct stars identified. The bottom panel reveals that most of the $\delta$\,Sct stars form a tight group, which could be interpreted as young stars immediately after the ZAMS that have just started fusing hydrogen. This ZAMS group, as we shall refer to it, is tighter in de-reddened colour than in TIC $T_{\rm eff}$, hence we continue to use colour in our analysis. We define the ZAMS group explicitly in Sec.\,\ref{ssec:pack}.

Within the ZAMS group, the small spread in brightness for a given colour is attributable to small age differences, or differences in rotation rate and the corresponding oblateness thus imposed. There is also a contribution from observational uncertainty: the pulsators have median uncertainties on $(G_{\rm BP} - G_{\rm RP})_0$ and $M_G$ of 0.051\,mag and 0.098\,mag, respectively, mostly originating from unknown extinction and associated reddening, although the uncertainties are dominated by systematic rather than random effects. The fact that the formal uncertainties in Fig.\,\ref{fig:CMD_box} are larger than the apparent scatter perhaps indicates that the extinction map uncertainties are overestimated.

Beyond the bright (upper) end of the ZAMS group lie many stars with high {\tt ruwe}, which are easily explained as binaries. There are many other stars that are around 1\,mag brighter than the ZAMS group, which do not have high {\tt ruwe} values. If their parallaxes are correct, these are probably not binaries, but Gaia is known to underestimate parallaxes (overestimate distances) to unresolved binaries in some cases \citep[e.g.][]{lee2021}, giving the impression they are much brighter than they are. It is probable that many of these stars do not  belong to Cep--Her. To avoid them affecting our statistics, we focus our analyses on the ZAMS group.


\subsection{Extracting mode frequencies}
\label{ssec:freqs}

For the $\delta$\,Sct stars, we determined mode frequencies via iterative prewhitening in a custom-written script,\footnote{\url{https://github.com/gautam-404/pre-whiten}} which iteratively fits sinusoids corresponding to peaks in the amplitude spectrum. Starting with the highest peak, the function fits for its frequency, amplitude, and phase, then subtracts it from the light curve. Iterations continue until a specified Signal-to-Noise Ratio (SNR) threshold is reached. The SNR is the ratio of the maximum amplitude in a given iteration to the median Fourier amplitude at that iteration.

Key thresholds and parameters incorporated within this module include:
\begin{itemize}
\item \texttt{snr\_threshold}: Dictates the SNR threshold for terminating iterations: if the $n$-th peak has an SNR below this threshold, iterations cease and the $n$-th peak is discarded. We used the default value of 5.

\item \texttt{nearby\_tolerance}: Specifies the permissible proximity between two frequencies before they are deemed overlapping. Ideally, i.e. in the case of a finite uninterrupted time-series of high signal-to-noise, this tolerance should be approximately the Rayleigh resolution ($1/T$), where $T$ is the observational time span. Our TESS light curves have T ranging from $\sim$27\,d to over 1000\,d, but for consistency we adopted a criterion of 0.01\,d$^{-1}$ as our {\tt nearby\_tolerance}. No peaks were sought at a frequency separation less than this from an existing peak.

\item \texttt{harmonic\_tolerance}: Determines the proximity between the frequency of a trial peak and an integer multiple of a different frequency, below which the trial peak would be deemed a harmonic. This tolerance is often set to a small frequency difference (typically smaller than $1/T$, e.g. \citealt{uzundagetal2023}), or a small fraction (say 1\%) of combination peak's frequency. This ensures that if, for example, one frequency is approximately twice (within 1\% deviation) another frequency, it is considered its harmonic and flagged as such. Since we were primarily interested in variability and not the relationship between peaks, we did not use this option.

\item \texttt{max\_iterations}: The maximum number of allowed iterations. We used the default value of 100.
\end{itemize}
The full documentation can be found in the \href{https://github.com/gautam-404/PreWhitener}{readme}.

We ran our pre-whitening code on all variable stars, as identified in Sec.\,\ref{ssec:variables}, using 10\,d$^{-1}$ as the lower frequency limit. The table of pulsation frequencies is available online (see Data Availability) and an excerpt is shown in Table\:\ref{tab:modes} for guidance on format and content.


\subsection{Measuring $\nu_{\rm max}$}
\label{ssec:numax}

By analogy with solar-like oscillations \citep{kjeldsen&bedding1995}, we use $\nu_{\rm max}$ to characterize the central frequency of the oscillations.  It is important to keep in mind that there is no single definition of $\nu_{\rm max}$, even for solar-like oscillations \citep{hekker2020, sreenivasetal2024}, and the situation is even more ambiguous for $\delta$\,Sct stars because of their uneven distribution of peak heights.
We measured $\nu_{\rm max}$ in a variety of ways to compare approaches and to ascertain a methodological uncertainty. The random uncertainty on each $\nu_{\rm max}$ method is very small, because the modes of $\delta$\,Sct stars are mostly coherent and of fairly constant amplitude \citep{murphyetal2014,bowmanetal2016}. However, the way in which $\nu_{\rm max}$ is determined can lead to differences of $\sim$10\:per\:cent. The simplest definition is to use the strongest peak, but $\delta$\,Sct stars can have modes excited at several radial orders \citep{antocietal2011,murphyetal2023}, and the strongest peak seldom lies at the centre of the envelope of excited peaks \citep[e.g.][]{beddingetal2023}. One can use a number of peaks to calculate an average, either in power or amplitude. 

We have measured $\nu_{\rm max}$ in five different ways. The first, which we refer to as $f_{\rm max}$, is simply the frequency of the strongest peak. Note that $f_{\rm max}$ is the same in amplitude and in power. We refer to the other four methods as `aggregate' methods because they use all significant peaks. These aggregate methods all used 10\,d$^{-1}$ as the lower frequency limit.

The second method calculates the moment of the $N$ extracted mode frequencies, $f_i$, weighted by their amplitudes, $A_i$ (measured in Sec.\,\ref{ssec:freqs}): 
\begin{eqnarray}
\nu_{\rm max,moment} = \frac{\Sigma_{i=1}^{N} f_i A_i}{\Sigma_{i=1}^{N} A_i}.
\end{eqnarray}
This is the method used by \citet{barcelofortezaetal2018}. The third method is identical to the second, except that it is calculated using power ($A_i^2$) instead of amplitude. The use of power is physically motivated, because the energy contained within a given mode is proportional to the power, rather than the amplitude. 

\begin{figure}
\begin{center}
\includegraphics[width=0.48\textwidth]{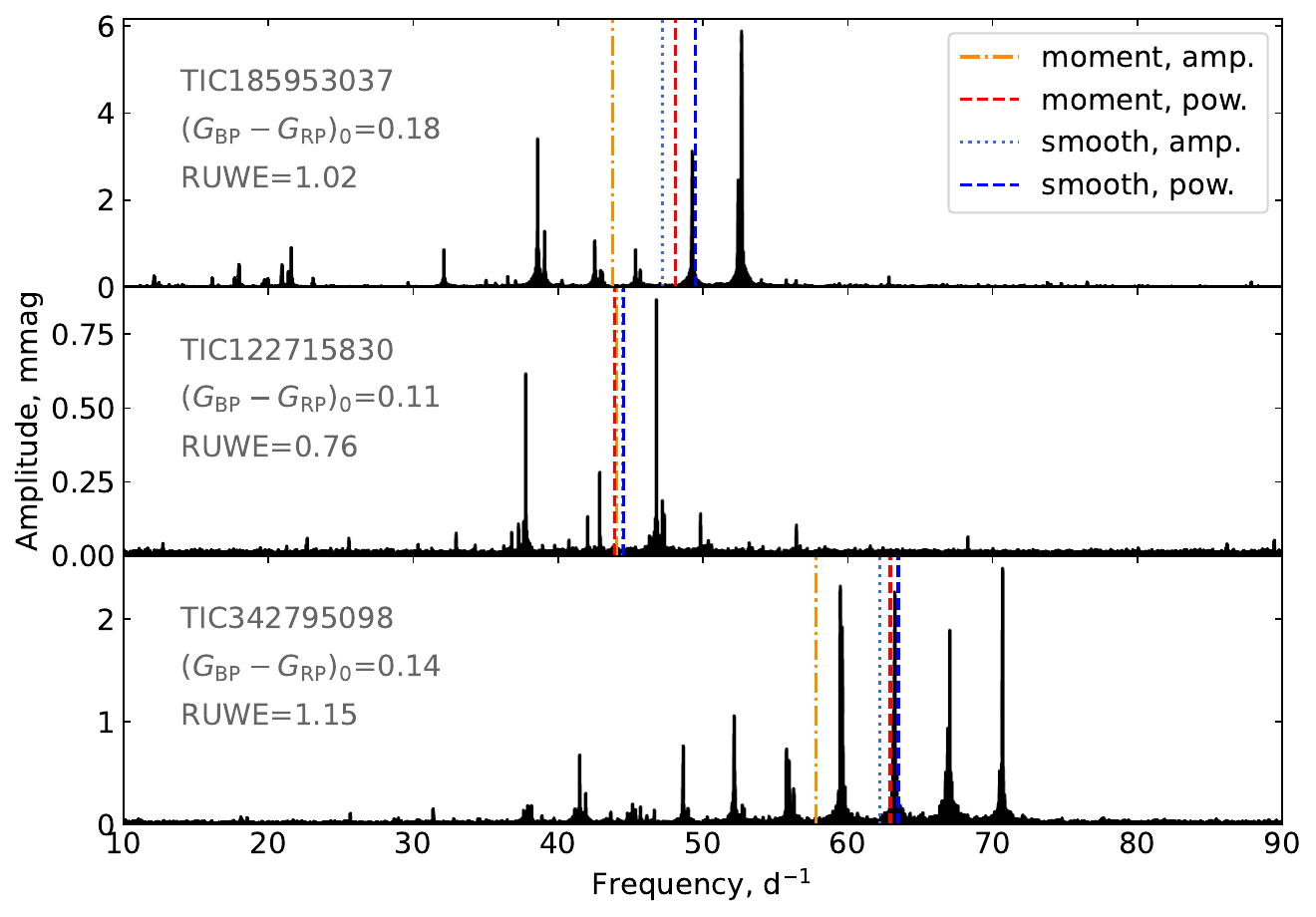}
\caption{Amplitude spectra of selected $\delta$\,Sct stars, showing $\nu_{\rm max}$ determinations with coloured lines (see legend). For some stars the methods produce a wide range of $\nu_{\rm max}$ values (top), whereas for other stars they agree well (middle). The methods determine $\nu_{\rm max}$ reliably when there are few peaks (middle) or many (bottom). All three targets were observed at 2-min cadence. The three stars are (top to bottom): HD\,192119, SAO\,68264 and HD\,348730.}
\label{fig:FT_numax}
\end{center}
\end{figure}

The fourth and fifth approaches were to heavily smooth the amplitude and power spectra by convolving with a broad Gaussian and measuring the peak of the resulting envelope. In studies of solar-like oscillations in main-sequence stars, a Gaussian of width $4\upDelta\nu$ is typically used \citep{kjeldsenetal2005,kjeldsenetal2008a}. Here, we used a width of 30\,d$^{-1}$, being about four times the $\upDelta\nu$ of the high-frequency $\delta$\,Sct stars in the association.\footnote{Convolution can be done with {\tt astropy.convolve\_fft} on a {\tt Gaussian1DKernel}. The {\tt fft} algorithm is much faster than {\tt convolve} for larger smoothing windows.} This approach requires the white noise to be subtracted first, lest the result be dominated by a spectrum of mostly noise in low-amplitude pulsators. For this, we measured the mean amplitude at each end of the Fourier data, $10<f<20$ and $f>70$\,d$^{-1}$, and took the lesser of these two values as the mean noise for each star separately. We subtracted this from each element of the amplitude array, allowing negative amplitudes.
Naturally, in both the moment and the smoothing method, the use of power will tend to bias the result towards the strongest peaks rather than a plethora of small ones. An example is provided in Fig.\,\ref{fig:FT_numax}.
 
Values of $\nu_{\rm max}$ resulting from all five methods are given in Table\,\ref{tab:big}. For some stars, all four aggregate methods agree well (as tightly as 0.06\,d$^{-1}$), whereas for others there is substantial spread (up to  13.76\,d$^{-1}$). The standard deviations of the aggregate methods around their collective mean are 1.90 (moment method in amplitude), 1.21 (moment method in power), 2.48 (smoothing method in amplitude), and 1.95\,d$^{-1}$ (smoothing method in power). (For the frequency of maximum amplitude, the standard deviation is 4.49\,d$^{-1}$.) We adopted the mean of the four aggregate methods hereafter.

\subsection{Defining the ZAMS group}
\label{ssec:pack}

\begin{figure}
\begin{center}
\includegraphics[width=0.48\textwidth]{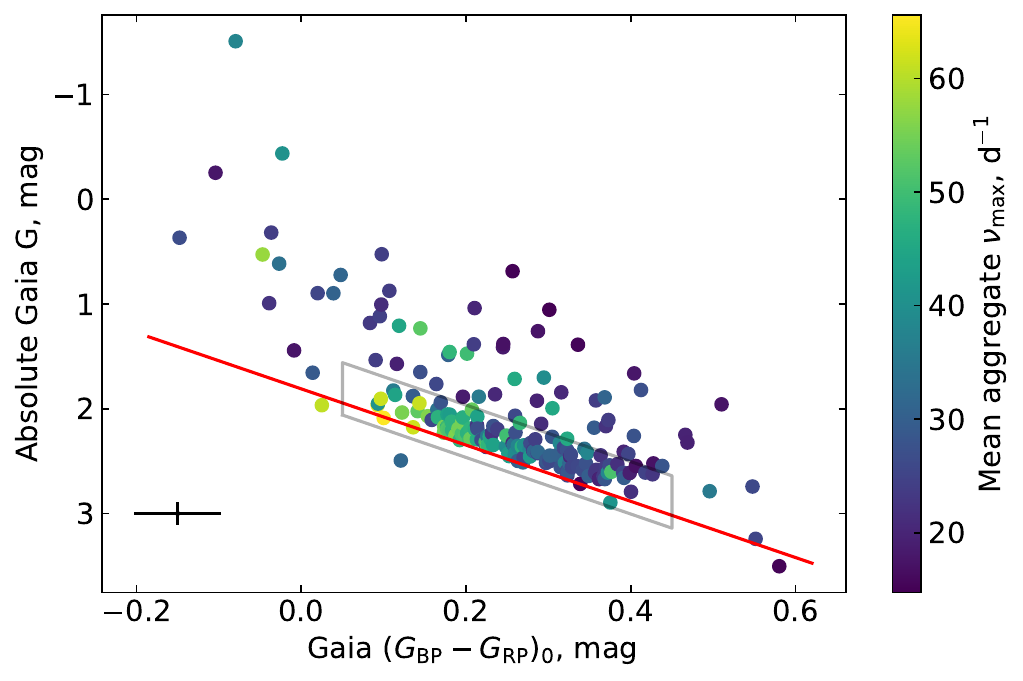}
\includegraphics[width=0.48\textwidth]{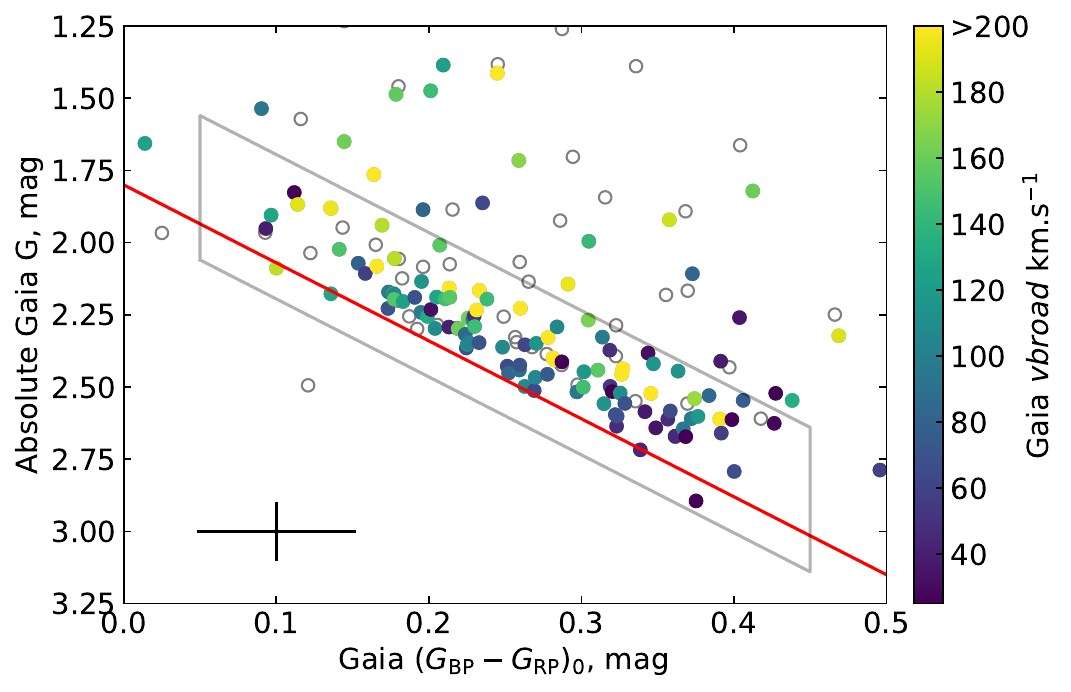}
\caption{Top: colour--magnitude diagram of the $\delta$\,Sct stars in the \mbox{Cep--Her} association, with plot symbols colour-coded by $\nu_{\rm max}$, as measured by the mean of the four aggregate methods. The location of the ZAMS is estimated with the red line. The grey box encloses the `ZAMS group' of coeval stars, which is described in the text and used for detailed analysis. Bottom: Zoom-in on the ZAMS group, showing the range of rotation velocities of the stars. Open grey symbols show stars without Gaia {\tt vbroad} data. The median uncertainty is shown as an error bar in the lower left corner of each panel. This is dominated by systematic rather than random uncertainty, hence the scatter of observations is somewhat smaller.}
\label{fig:CMD_box}
\end{center}
\end{figure}

\begin{figure}
\begin{center}
\includegraphics[width=0.48\textwidth]{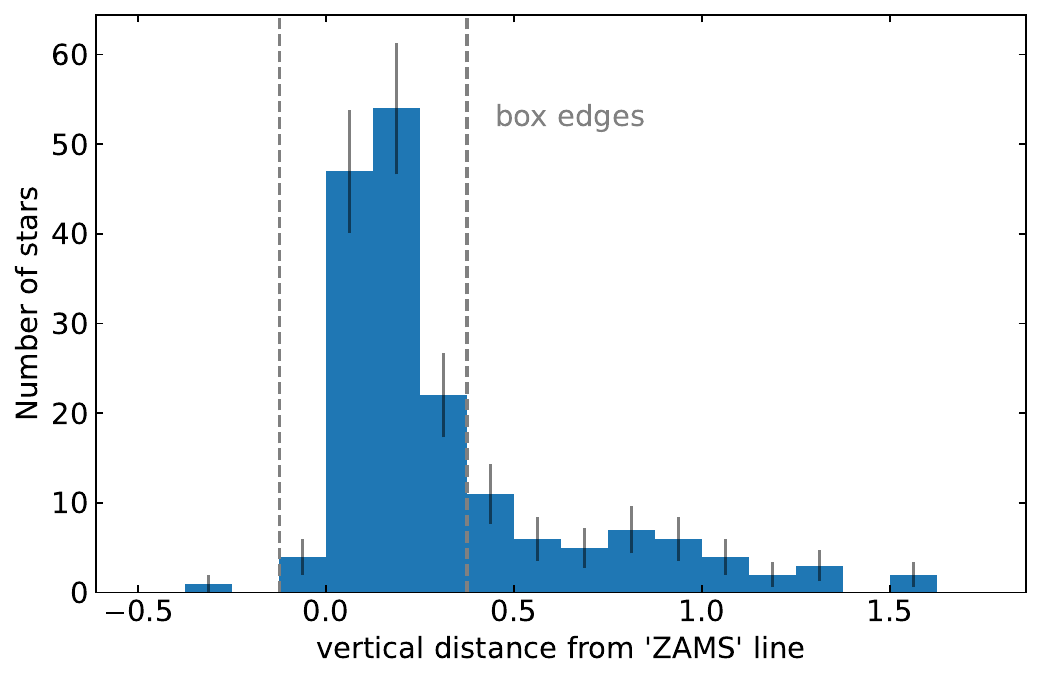}
\caption{A histogram of the vertical distance from the ZAMS line in Fig.\,\ref{fig:CMD_box}.}
\label{fig:vert_hist}
\end{center}
\end{figure}

In Fig.\,\ref{fig:CMD_box}  we show $\nu_{\rm max}$ as measured by the moment (amplitude) method across the colour--magnitude diagram. Amongst the ZAMS group, but not outside it, there is an apparent trend of higher $\nu_{\rm max}$ at bluer temperatures, that is, an apparent $\nu_{\rm max}$--$T_{\rm eff}$ relation. We defined the ZAMS group numerically for further analysis, based on their height above the ZAMS, as follows.
We first defined a line that roughly locates the ZAMS, by tracing the minimum luminosity of the ZAMS group (red line in Fig.\,\ref{fig:CMD_box}), which has the equation
\begin{eqnarray}
M_G = 2.7 (G_{\rm BP} - G_{\rm RP})_0 + 1.8.
\end{eqnarray}
The grey box surrounding it constrains the sample to colours of $0.05<(G_{\rm BP} - G_{\rm RP})_0<0.45$, which emerged as the natural edge of the population. The vertical height of the box is 0.5\,mag, and extends 0.125\,mag below the ZAMS to capture slightly fainter stars, such as those that might be partially obscured by circumstellar disks or might have been shifted down by larger-than-average parallax uncertainties. The choice of box height excluded several binaries identified by their {\tt ruwe} values in Sec.\,\ref{ssec:variables}. The histogram in Fig.\,\ref{fig:vert_hist} shows the vertical distance of stars from the ZAMS line and justifies the edge of the ZAMS group drawn in this manner: above the drawn box, the number of stars per bin falls to a background level, within the uncertainties. We also note the clear tendency of rapid rotators in Fig.\,\ref{fig:CMD_box} to lie farther above the ZAMS line than the slow rotators, as expected. The parameters of the 126 ZAMS-group $\delta$\,Sct stars are given in Table\:\,\ref{tab:big}. We investigate their $\nu_{\rm max}$--$T_{\rm eff}$ relation in the next section.

\section{The $\nu_{\rm\lowercase{max}}$--$T_{\rm \lowercase{eff}}$ scaling relation}
\label{sec:scaling}

Solar-like oscillators exhibit a power excess in a roughly Gaussian envelope, inside which all eigenmodes are excited.
The observational challenge is to obtain a time-series of sufficient precision, duration and cadence to see them. The centre of this envelope is called $\nu_{\rm max}$ and follows one of the asteroseismic scaling relations \citep{brownetal1991,kjeldsen&bedding1995},
\begin{eqnarray}
\nu_{\rm max} \propto g/ \sqrt{T_{\rm eff}},\label{eq:numax}
\end{eqnarray}
where $g$ is the surface gravity (typically measured in solar units). Main-sequence stars, subgiants, and red giants follow this relation quite closely \citep[e.g.][]{chaplin&miglio2013,hekker2020, ylietal2021}.

The $\delta$\,Sct stars, on the other hand, typically have complex and seemingly disordered amplitude spectra that are not usually contained within a roughly Gaussian envelope \citep[e.g.][]{guzik2021,ramon-ballestaetal2021}. In fact, it only appears to be young $\delta$\,Sct stars that provide an exception to this, where comb-like patterns of pulsation modes are sometimes seen \citep{ suarezetal2014, paparoetal2016, micheletal2017, beddingetal2020, murphyetal2021a, murphyetal2022, steindletal2022, beddingetal2023, scuttetal2023, murphyetal2023}. As detailed by \citet{murphyetal2023}, the reasons for this are becoming understood, which are principally that older stars feature greater interaction between the p\: and g\:modes \citep{christensen-dalsgaard2000,lignieres&georgeot2009,aertsetal2010} and older stars have sharp molecular weight gradients at the edge of the convective core \citep{reeseetal2017,dornan&lovekin2022}, both of which spoil regular patterns. But even in these young $\delta$\,Sct stars, a true envelope is seldom apparent: \citet{beddingetal2023} showed that the distribution of peaks in the amplitude spectra of $\delta$\,Sct stars in the Pleiades cluster is not very ordered, even though a correlation is expected between photometric colour (or $T_{\rm eff}$) and the excitation of pulsation modes of higher radial orders \citep{dziembowski1997,pamyatnykh2000}.

Conversely, studies of large numbers of $\delta$\,Sct stars do find the expected correlation between $\nu_{\rm max}$ and $T_{\rm eff}$ \citep{bowman&kurtz2018,barcelofortezaetal2018,barcelofortezaetal2020,hasanzadehetal2021}. In the open cluster NGC\,2516, which has an age of about 100\,Myr, \citet{glietal2024} found a clear relation between pulsation frequency and colour for a subset of stars in a narrow colour range. We have already shown that there is a correlation in the  \mbox{Cep--Her} sample in Sec.\,\ref{ssec:pack}. In this section, we investigate whether differences in method (i.e. systematic uncertainties) might play a role in hiding the correlation, and whether the correlation is useful for estimating stellar properties.


\subsection{Systematic uncertainties in measuring $\nu_{\rm max}$}
\label{ssec:systematics}

By analysing only stars in the ZAMS group (Sec.\,\ref{ssec:variables} and \ref{ssec:pack}), we have a homogeneous sample in terms of metallicity, surface gravity, and age (within a few tens of Myr). Our sample selection will have preferentially excluded binaries, especially those in which the $\delta$\,Sct star is the secondary. Hence, any binaries that remain in the sample should have similar colours to the $\delta$\,Sct component they contain, and should not strongly affect any $\nu_{\rm max}$--$T_{\rm eff}$ relation. 

\begin{figure}
\begin{center}
\includegraphics[width=0.48\textwidth]{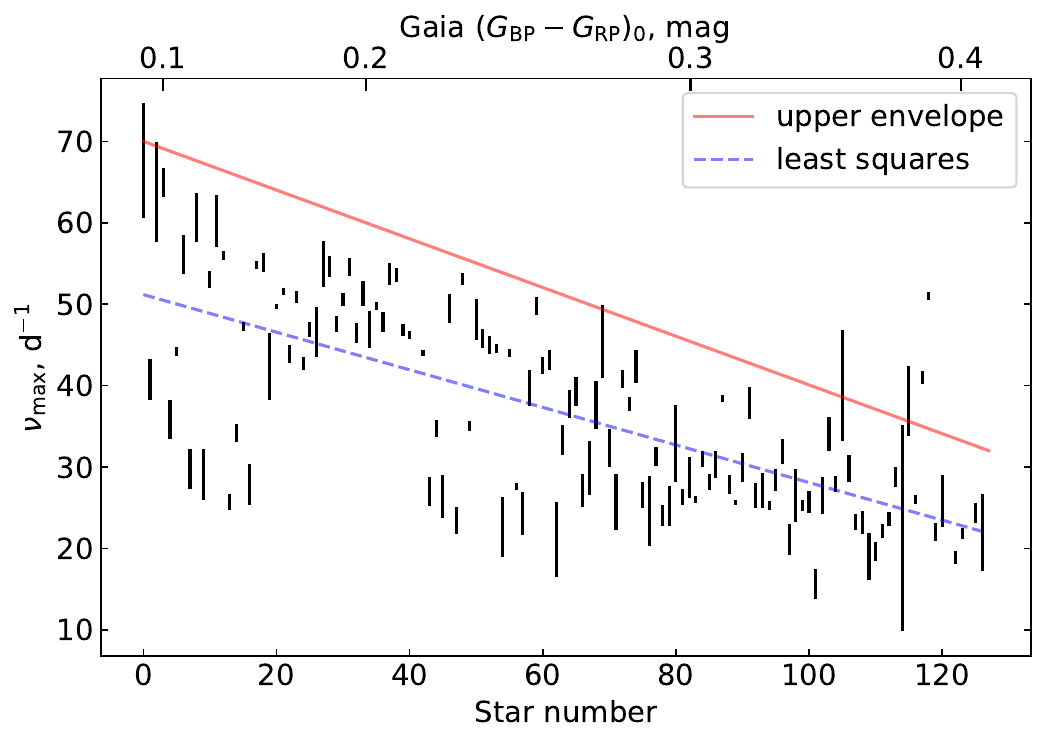}\\
\includegraphics[width=0.48\textwidth]{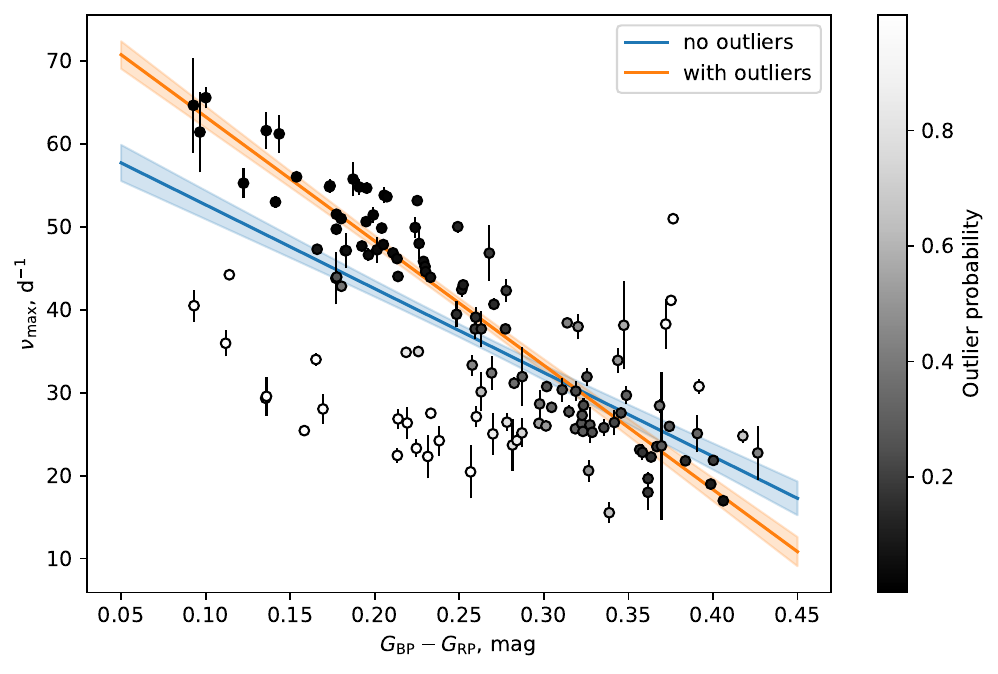}\\
\includegraphics[width=0.48\textwidth]{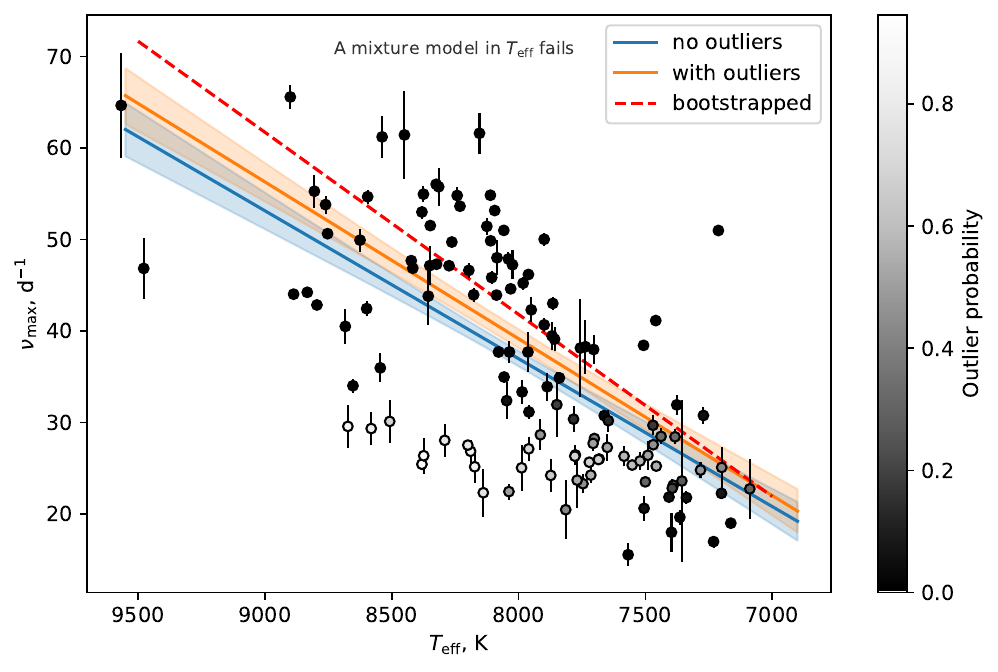}\\
\caption{Fitting a $\nu_{\rm max}$--$T_{\rm eff}$ relation for the ZAMS group of stars in Cep--Her. Top: Black lines show the range of $\nu_{\rm max}$ values determined by the four methods for stars in the ZAMS group, sorted by colour. A least-squares fit using each star's mean $\nu_{\rm max}$ is shown in blue, and an apparent upper envelope is highlighted in red for discussion (Sec.\,\ref{ssec:systematics}). Middle: A mixture model of the sample, in which stars either belong to the relation (low outlier probability; dark points) or belong to a Gaussian background population instead (high outlier probability; light points). Two linear fits are shown: the outlier model in orange, and the linear fit to all points (no outlier model) as per the top panel in blue. Bottom: as for the middle panel with the ordinate axis changed to $T_{\rm eff}$, showing that a Gaussian outlier model only works for this sample in colour space. The dashed red line is a linear fit based on outlier probabilities from the middle panel (see text).}
\label{fig:lines}
\end{center}
\end{figure}

In Fig.\,\ref{fig:lines} we show our four aggregate methods (Sec.\,\ref{ssec:numax}) for measuring $\nu_{\rm max}$ (omitting the measurement that uses only a single peak) as a function of $(G_{\rm BP} - G_{\rm RP})_0$ colour. The figure shows an upper envelope: high $\nu_{\rm max}$ values are not found for redder stars, though some blue stars do have low $\nu_{\rm max}$. The linear least-squares fit in the top panel of Fig.\,\ref{fig:lines} is clearly a poor fit to the data. This cannot be explained by sample inhomogeneity, or by measurement uncertainty, because we have been generous in applying different methodologies for measuring $\nu_{\rm max}$ and none of these methods has large random uncertainties. At face value, it seems the relation is weak.

In the middle panel of Fig.\,\ref{fig:lines} we take a more statistical approach. We assume that there is a $\nu_{\rm max}$--$T_{\rm eff}$ relation that is obscured by `outlier' stars that for some unknown reason do not fit the relation. Thus, we generated a mixture model of two populations: those that fit the relation, and those that do not. We used $(G_{\rm BP} - G_{\rm RP})_0$ colour as the independent variable, and we used the mean and standard deviation of the four aggregate methods for the $y$ data and their uncertainties, respectively.\footnote{For the purpose of minimising uncertainty in the modelling process, the data were reparametrized to have a mean of zero in both $x$ and $y$, and were transformed back afterwards.} The mixture model presumed some stars to follow a $\nu_{\rm max}$--$T_{\rm eff}$ relation modelled with a linear fit whose slope and intercept were free parameters. The background population was modelled as a colour-independent Gaussian in $\nu_{\rm max}$ whose mean and standard deviation were free parameters. We also added a jitter term to account for the scatter of points around the relation, which is much larger than the error bars on the data. This was preferred over rescaling the error bars because visual inspection showed that the $\nu_{\rm max}$ values and their uncertainties were well determined (see Fig.\,\ref{fig:FT_numax}). Finally, we added one additional variable which was the probability of each point being an outlier. We ran this mixture model using a No U-Turns Sampler (NUTS) in {\tt numpyro} with {\tt jax}. The sampler used two chains, had $\sim$3500 effective samples, and we checked for convergence with the Gelman--Rubin (\^r) statistic using the {\tt arviz} package. The resulting $\nu_{\rm max}$-colour relation is 
\begin{eqnarray}
\nu_{\rm max}~({\rm d}^{-1}) = -149.8~[(G_{\rm BP} - G_{\rm RP})_0/{\rm mag}] + 78.2, \label{eq:bmr}
\end{eqnarray}
which is shown as the orange line in the middle panel of Fig.\,\ref{fig:lines}.

We note that the same model cannot be applied when the independent variable is $T_{\rm eff}$ instead of colour (Fig.\,\ref{fig:lines}, bottom) because the distribution of `outlier' stars maps across differently. Specifically, we found that in $T_{\rm eff}$-space a Gaussian was not a good representation of a background population -- it becomes too heavily skewed towards the group of stars with $\nu_{\rm max}$ near 25\,d$^{-1}$. Since there is no physically-motivated reason to prefer some other functional form for `outliers', we used the middle panel of Fig.\,\ref{fig:lines} to identify stars with $>$85\% probability of lying on the relation, and used their TIC $T_{\rm eff}$ values to calculate a $\nu_{\rm max}$--$T_{\rm eff}$ relation. From this we found
\begin{eqnarray}
\nu_{\rm max}~({\rm d}^{-1}) =  0.0203~(T_{\rm eff}/{\rm K}) -120.2, \label{eq:teff}
\end{eqnarray}
which is shown as the `bootstrapped' line in the bottom panel of Fig.\,\ref{fig:lines}. Note that we did not include a surface gravity term for two reasons. First, unlike solar-like oscillators, $\delta$\,Sct stars do not pulsate at frequencies near the acoustic cut-off frequency, and the reason for their pulsation frequencies to depend on $T_{\rm eff}$ is entirely different from solar-like oscillators. 
Second, our sample is so tightly confined in the colour--magnitude diagram already that no $\log g$ data would be attainable with sufficient accuracy and precision to tighten this relation further.

The origin of the temperature-independent `outliers' remains unclear. We inspected their amplitude spectra to determine whether they were unusual, and to reaffirm that the automated methods to determine $\nu_{\rm max}$ had worked correctly. Specifically, we considered the stars with \mbox{$(G_{\rm BP} - G_{\rm RP})_0<0.2$} that have a mean $\nu_{\rm max}<50$\,d$^{-1}$. The outliers were unremarkable, and the $\nu_{\rm max}$ methods had worked as intended. The examples we showed in Fig.\,\ref{fig:FT_numax} illustrate this well. The top and middle panels show stars that might be considered outliers: they have \mbox{$(G_{\rm BP} - G_{\rm RP})_0<0.2$} and $\nu_{\rm max}<50$\,d$^{-1}$, but look like ordinary $\delta$\,Sct stars. One of these (middle panel) has all four methods agreeing on the $\nu_{\rm max}$ value, whereas the other (top panel) has a wider range of values, but all are reasonable approximations to the centre of the variability. Contrasting the star in the bottom panel of Fig.\,\ref{fig:FT_numax} with the one in the middle panel, we see two stars with low {\tt ruwe} (presumably single stars) with similar colours (0.14 vs 0.11 mag) but vastly different amplitude spectra.


 \subsection{Is the $\nu_{\rm max}$--$T_{\rm eff}$ relation useful?}
 \label{ssec:useful}
 
In the TESS era where most $\delta$\,Sct stars have light curves available, either from FFIs or otherwise, it would be convenient to be able to use the amplitude spectrum to estimate $T_{\rm eff}$. This would obviate some of the issues with using colour as a $T_{\rm eff}$ proxy, such as unknown reddening, and might provide tighter and/or more accurate constraints on asteroseismic modelling. This approach has been used to model $\delta$\,Sct stars in the open clusters $\alpha$\,Per \citep{pamosortegaetal2022}, Trumpler~10 and Praesepe \citep{pamosortegaetal2023}. However, we can see that even for stars in the Cep--Her ZAMS group, which represent a homogeneous sample, there can be large outliers in $\nu_{\rm max}$, and stars lying just outside the ZAMS group present an even bigger problem (contrast the $\nu_{\rm max}$ of stars inside with those just above the box in Fig.\,\ref{fig:CMD_box}).
 
The standard deviation of points around the $\nu_{\rm max}$--$T_{\rm eff}$ relation of eq.\,\ref{eq:teff} when outliers are excluded is 372\,K. (Equivalently, in $\nu_{\rm max}$ it is 7.4\,d$^{-1}$.) At 8000\,K, which is the median of the sample, this is a 4.7\% uncertainty. By comparison, the $T_{\rm eff}$ scale is calibrated by interferometry at the 2\% level (\citealt{casagrandeetal2014,whiteetal2018}; cf. \citealt{milleretal2020,milleretal2022}), and spectroscopic $T_{\rm eff}$ uncertainties for A stars are typically around 250\,K, i.e. $\sim$3\:per\:cent, or smaller \citep[e.g.][]{niemczuraetal2017,kahramanalicavusetal2020}. In this sense, the relation in eq.\,\ref{eq:teff} might offer a useful approximation when no better alternative exists. However, the much larger concern is the presence of the temperature-independent `outlier' population. It is not possible to know to which population a random field star belongs.

Moreover, the 372-K uncertainty corresponded to our careful analysis with robust `outlier' treatment. Suppose that one were to treat the ZAMS group stars as a single population instead, fitting a single linear relation like the dashed-blue line in the top panel of Fig.\,\ref{fig:lines}. The scatter (1$\sigma$ standard deviation) would then be 705\,K. And if {\it all} stars in our initial Cep--Her membership list with $P_{\rm mem}>0.25$ and with  between $7000 < T_{\rm eff} < 10$\,000\,K are fitted instead (i.e. temporarily disbanding the ZAMS group), the scatter would equal $\pm$1065\,K. The instability strip itself is only 2000\,K wide \citep{dupretetal2005b,murphyetal2019}. Hence, we are forced to conclude that although the $\nu_{\rm max}$ and $T_{\rm eff}$ of {\it some} $\delta$\,Sct stars in Cep--Her are correlated, there is no useful relation between them.


\subsection{The $\nu_{\rm max}$--$T_{\rm eff}$ relation in other clusters}
\label{ssec:multicluster}

A handful of clusters have now been studied for evidence of a $\nu_{\rm max}$--$T_{\rm eff}$ relation. Most of these are the same age: Cep--Her spans ages 25--80\,Myr, Trumpler-10 has an age of $\sim$50\,Myr \citep{kerretal2023}, NGC\,2516 is around 100-Myr old \citep{glietal2024}, and the Pleiades cluster has an age of roughly 125\,Myr. Praesepe has also attracted attention, but is substantially older at several hundred Myr. We compare their members' $\nu_{\rm max}$ and $T_{\rm eff}$ values here.

\begin{figure}
\begin{center}
\includegraphics[width=0.48\textwidth]{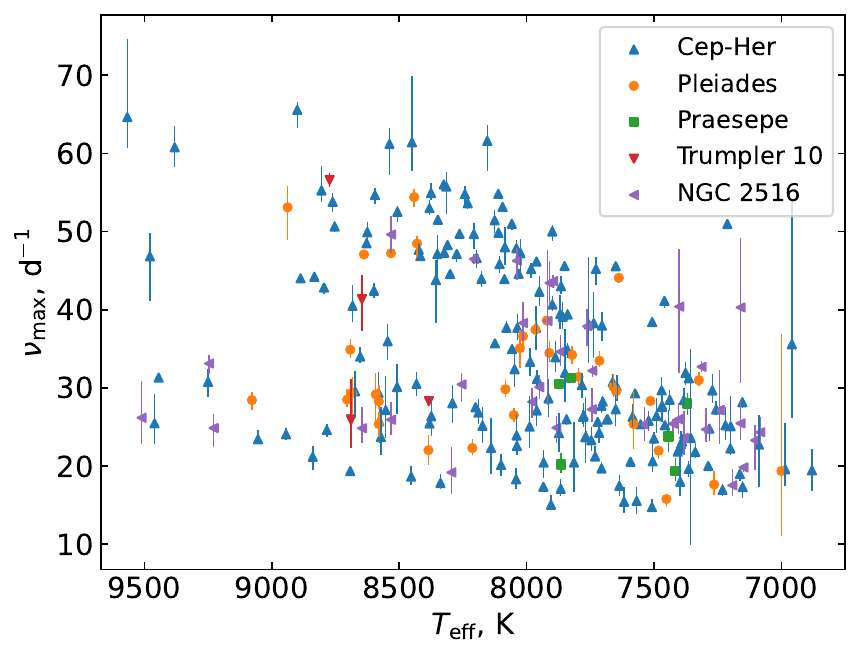}\\
\caption{The distributions of $\nu_{\rm max}$ and $T_{\rm eff}$ for the handful of clusters described in Sec.\,\ref{ssec:multicluster}. Symbols denote the mean of the aggregate methods for measuring $\nu_{\rm max}$, while vertical bars span the range of $\nu_{\rm max}$ from those methods for each star. One of the Trumpler~10 stars lies outside the plot range (TIC\,30307085, $\nu_{\rm max} = 61.0$\,d$^{-1}$, $T_{\rm eff} = 9931$\,K), but consistency with Fig.\,\ref{fig:lines} was retained. Of interest is the emerging structure of two ridges that merge around 8000\,K.}
\label{fig:multicluster}
\end{center}
\end{figure}

Data were obtained and analysed as uniformly as possible, in the same manner as for Cep--Her. Specifically, we gathered TIC numbers of $\delta$\,Sct members and corresponding TIC temperatures from the relevant studies, namely, for 36 stars in the Pleiades from \citet{beddingetal2023}, for 33 stars in NGC\,2516 from \citet{glietal2024}, and for both 6 stars in Praesepe and 5 stars in Trumpler~10 from \citet{pamosortegaetal2023}. With the exception of NGC\,2516, for which we requested the custom FFI light curves from the authors, we downloaded 2-min and 10-min TESS light curves in that order of preference as we did for Cep--Her. The four aggregate methods for measuring $\nu_{\rm max}$ were applied identically across all clusters. We show the distributions of $\nu_{\rm max}$ and TIC $T_{\rm eff}$ in Fig.\,\ref{fig:multicluster}.

Unlike in Fig.\,\ref{fig:lines}, where we focussed on only those stars in the ZAMS group, here all Cep--Her stars in the shown $T_{\rm eff}$ range are plotted, since the membership lists of other clusters were not curated in the same way as the ZAMS group. The $\nu_{\rm max}$--$T_{\rm eff}$ relation is shown in Fig.\,\ref{fig:multicluster} and some interesting structure is evident. Specifically, very few stars lie in the upper-right portion of the diagram, and stars hotter than $T_{\rm eff}\sim8000$\,K appear to fall on either one of a pair of (non-parallel) ridges.

Briefly, we might explain this as follows: for young clusters like those in Fig.\,\ref{fig:multicluster}, the typical $\Delta\nu$ is around 7\,d$^{-1}$ (we show this in Sec.\,\ref{ssec:rot-puls} for Cep--Her stars), and the fundamental radial mode frequency lies at roughly 3$\Delta\nu$ \citep{murphyetal2023}, hence the horizontal (temperature-independent) band of points comprises stars dominated by the fundamental mode. Although the figure covers stars of a wide range of masses, it appears that ZAMS densities of $\delta$\,Sct stars are mass-independent \citep{murphyetal2023}, unless modified by rotation, hence their $\Delta\nu$ values and fundamental mode frequencies are similar across a range of temperatures (masses). Since the fundamental radial mode is the p\:mode of lowest frequency, modes of the same radial order but different degree as well as modes at the next radial order ($n=2$) will have slightly higher frequencies, which explains why this ridge in the figure constitutes a broadened band between 20 and 30\,d$^{-1}$. The stars that lie on the diagonal band in Fig.\,\ref{fig:multicluster}, whose $\nu_{\rm max}$ exhibits temperature dependence, would necessarily oscillate in modes of higher radial order.

The question is then why do some hot $\delta$\,Sct stars pulsate predominantly in low radial orders (around the fundamental radial mode) while others pulsate at higher orders? The same question, incidentally, might be applied to the so-called second ridge in the Period--Luminosity relation \citep{ziaalietal2019,baracetal2022}. One can speculate that the answer is connected to the different driving mechanisms behind $\delta$\,Sct pulsations (\citealt{houdek2000,antocietal2014,murphyetal2020d}), but that only shifts the question to why the same available mechanisms drive oscillations of higher radial orders in some stars of the same temperature than in others.

An answer might be found in studies of roAp stars, for which models whose surface layers are depleted in helium have unstable p\:modes at higher frequencies than those without helium depletion \citep{cunhaetal2013}. In other words, the level of helium depletion affects $\nu_{\rm max}$. Suppose this also applies to $\delta$\,Sct stars without the (typically strong) magnetic fields seen in roAp stars, which appears to be the case (M. Cunha, private communication). In that case, two predictions follow: (i) unless a substantial amount of helium settling can occur soon after the fully-convective pre-MS phase is over, then older stars (i.e. older clusters) should behave differently from younger ones; and (ii) rapid rotators should behave differently from slow rotators, since helium settling is ineffective in the former due to large scale interior flows (meridional circulation). 

The first prediction is unfortunately not addressed by the Praesepe members in Fig.\,\ref{fig:multicluster}. Praesepe is much older than the other clusters but its $\delta$\,Sct members are all cooler than 8000\,K, hence they do not lie in the part of the diagram where we see two ridges. Another intermediate-age cluster (0.2--0.5\,Gyr) is required to evaluate this hypothesis, noting that older clusters will have fewer hot $\delta$\,Sct stars (they evolve to cooler $T_{\rm eff}$). We could find none with readily available data, and a full analysis of another cluster is beyond the current scope.

\begin{figure}
\begin{center}
\includegraphics[width=0.48\textwidth]{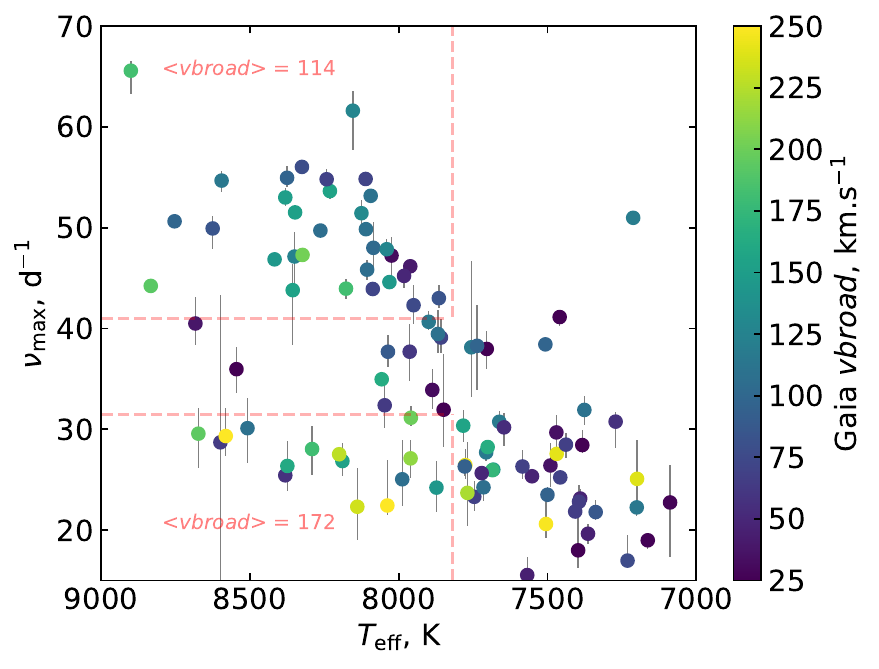}\\
\caption{The $\nu_{\rm max}$ and $T_{\rm eff}$ values for Cep--Her stars in the ZAMS group, colour-coded by their Gaia {\tt vbroad} values. Dashed lines separate stars that are unambiguously on the bottom ridge from those on the upper ridge, and the mean {\tt vbroad} within each division is shown.}
\label{fig:vbroad_numax}
\end{center}
\end{figure}

The second prediction can be evaluated with our current data. Using the Cep--Her stars in the ZAMS group that have Gaia {\tt vbroad} measurements, we observe some dependence on {\tt vbroad} (Fig.\,\ref{fig:vbroad_numax}). Specifically, the most rapid rotators appear to lie on the bottom ridge. By dividing the plot into areas containing stars that are unambiguously bottom-ridge or top-ridge stars, we see substantial differences in their mean {\tt vbroad} values, at 172 and 114\,km\,s$^{-1}$, respectively. The standard deviation around the former mean is 65\,km\,s$^{-1}$, hence this is only a 1$\sigma$ result. We can say only that there is weak evidence that rapid rotation causes $\delta$\,Sct stars to pulsate in low-$n$ modes.

\section{The $\delta$\,Sct pulsator fraction in Cep--Her}
\label{sec:frac}

\begin{figure}
\begin{center}
\includegraphics[width=0.48\textwidth]{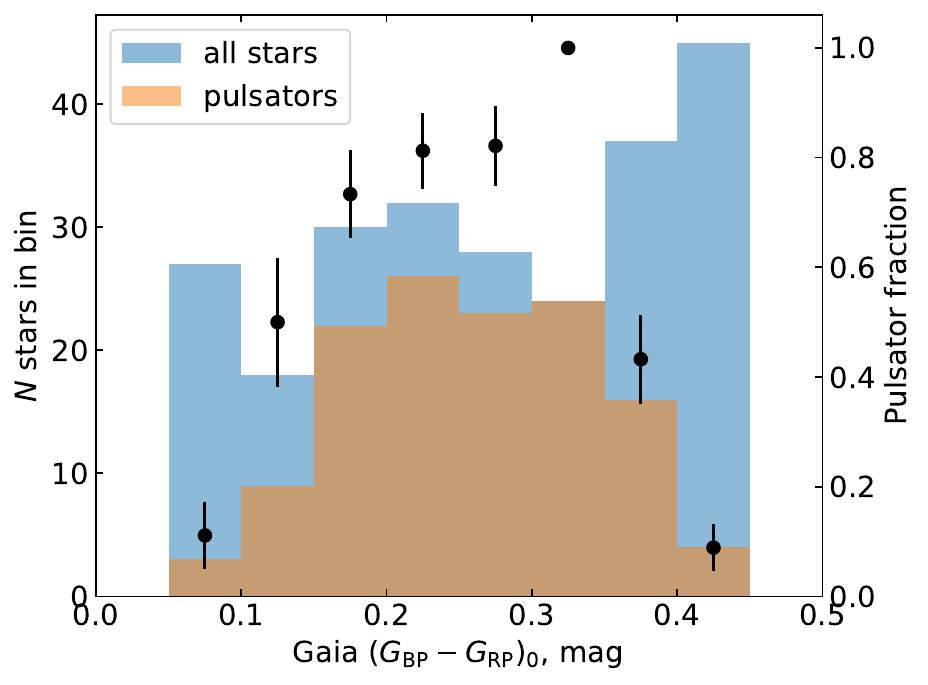}
\caption{Histograms (left $y$-axis) of the 241 stars in the ZAMS group, and the 127 of those that pulsate, as a function of colour. Black circles show the pulsator fraction (right $y$-axis), with uncertainties calculated via Eq.\,\ref{eq:frac_err}.}
\label{fig:hist}
\end{center}
\end{figure}

Using the `dSct' flag assigned in Sec.\,\ref{ssec:variables}, and evaluating stars within the box containing the ZAMS group on the HRD (Sec.\,\ref{ssec:pack}), we calculated a histogram of the number of pulsators and of the total number of stars ($N$) as a function of dereddened $(G_{\rm BP} - G_{\rm RP})_0$ colour (Fig.\,\ref{fig:hist}). We also show the pulsator fraction, $f$, as the quotient of those quantities. The uncertainty on that quotient is calculated for the $i$-th bin by assuming the measurements comprise a set of binomial trials (also called Bernoulli trials), in which the star is either pulsating or it is not (e.g. \citealt{taylor2022}):
\begin{eqnarray}
\sigma_{f,i} = \sqrt\frac{f_i(1-f_i)}{N_i}. \label{eq:frac_err}
\end{eqnarray}
We found that the pulsator fraction rises monotonically to its peak in the bin \mbox{$0.30 \leq (G_{\rm BP} - G_{\rm RP})_0 \leq 0.35$}, where it reaches 100\%. In $T_{\rm eff}$, this bin corresponds to $7750 \gtrsim T_{\rm eff} \gtrsim 7500$\,K.

The emerging trend is that younger populations have a higher fraction of $\delta$\,Sct stars. The \textit{Kepler} $\delta$\,Sct population from \citet{murphyetal2019} comprised over 2000 stars that were slightly out of the Galactic plane, and the distance from the Galactic plane was generally larger for the more distant stars. Their average age is presumably several hundred Myr. \citet{murphyetal2019} measured their pulsator fraction as 50--60\:per\:cent in the middle of the instability strip. The $\sim$120-Myr Pleiades sample from \citet{beddingetal2023} was much smaller (36 $\delta$\,Sct stars) and yielded a markedly higher pulsator fraction: $84\pm7$\:per\:cent in middle of the instability strip [$0.2<(G_{\rm BP} - G_{\rm RP})_0<0.4$]. While the Cep--Her sample is not as well characterised as the Pleiades, it is larger. There are as many Cep--Her $\delta$\,Sct stars in individual 0.05-mag bins as in the entire Pleiades sample. It shows a temperature dependence in the pulsator fraction, as did the \textit{Kepler} sample, and it peaks at 100\:per\:cent ($N$=25). This is the first detection of a `pure' $\delta$\,Sct instability strip in any sample. The result is hardly an artefact of binning: shifting the bin edges by half a bin results in 24 of 25 stars being pulsators for a pulsator fraction of $96\pm4$\:per\:cent. The age range of the stars in this diverse star complex (25--80\,Myr) is broader and systematically younger than the Pleiades.

Why does the pulsator fraction decrease with age? The answer must depend on the ratio of pulsational driving and damping, and how this changes as the star evolves. The evolution itself influences both: as A-type stars evolve, their radii increase and their surfaces become cooler. There is stronger convection at the surface, and the He\,{\sc ii} partial ionization zone, which is the primary driver of pulsations via the $\kappa$~mechanism, moves deeper in the star. However, these changes in driving and damping are not necessarily the strongest factors. As foreshadowed in Sec.\,\ref{ssec:multicluster}, a more important factor appears to be the fact that in slow rotators helium gravitationally sinks out of the driving zone \citep{baglinetal1973}. The lack of rotational mixing \citep[e.g.][]{michaudetal1983} also manifests as deeper metal lines in the spectra \citep{titus&morgan1940}, from elements radiatively levitated towards the surface \citep[e.g.][]{theadoetal2009,dealetal2020}. These are the metallic-lined (Am) stars, and they appear to have a lower pulsator fraction than chemically normal stars \citep{durfeldt-pedrosetal2024}. 
Importantly, the onset of helium depletion is quite fast: even though helium continues to deplete across 1000\,Myr in a 1.6-M$_{\odot}$ star, 50\% of that depletion has occurred by 100\,Myr and 80\% has occurred by 500\,Myr \citep{theadoetal2005}.\footnote{The \citet{theadoetal2005} models include magnetic fields, but a similar result is found by \citet{dealetal2016} without magnetic fields and including more microscopic processes such as thermohaline convection, which can have macroscopic effects. The latter study quotes only the mean molecular weight, rather than helium abundance directly, hence the former study is more readily interpretable.} By comparison, the detectable accumulation of heavier elements, such as iron, near the surface seems to take a little longer, having time-scales of hundreds of Myr \citep{dealetal2016}. The time-scales for helium depletion are consistent with the observed pattern of a higher pulsator fraction in \mbox{Cep--Her} (and in the Pleiades) than in the Kepler sample.

Similarly relevant are the Ap and the $\lambda$\,Boo stars. Convective mixing is suppressed by strong global---and, presumably, dipolar---magnetic fields in the Ap stars \citep{theado&cunha2006}. This not only facilitates helium sinking from the He\,{\sc ii} driving zone but also changes the work integrals, resulting in suppression of $\delta$\,Sct-like pressure modes \citep{saio2005}. Yet most observed Ap stars are old, suggesting that these processes take some time, and young stars might be scarcely affected. In other words, the mechanisms producing the peculiarities of both the Am and Ap stars are expected to preferentially reduce the pulsator fraction of older stars, even if neither suppresses pressure-modes completely \citep{smalleyetal2014,murphyetal2020d}. The $\lambda$\,Boo stars show the opposite effect. They are hypothesized to have extra helium in their He\,{\sc ii} partial ionization zones because of recent accretion of circumstellar gas \citep{kamaetal2015}, and they tend to have young ages \citep{folsometal2012,murphy&paunzen2017}. The pulsator fraction of $\lambda$\,Boo stars has been measured to be significantly higher than for normal stars \citep{murphyetal2020b}, which is consistent with our observations here that younger stars have a higher pulsator fraction. Indeed, one might expect many of the $\delta$\,Sct stars in \mbox{Cep--Her} to have the $\lambda$\,Boo abundance pattern.

Also relevant to this discussion is the all-sky sample of TESS $\delta$\,Sct stars in the middle of the instability strip ($0.29< G_{\rm BP} - G_{\rm RP} <0.31$, not de-reddened) studied by \citet{readetal2024}. This population spans a range of masses, ages, and metallicities like the \textit{Kepler} sample did. Still, its stars have brighter apparent magnitudes on average, so they are generally nearby and hence confined to the Galactic thin disk. They should be slightly younger than the average star in the \textit{Kepler} sample. The pulsator fraction of the \citet{readetal2024} stars peaks at $70\pm10$\%, i.e. at a slightly higher fraction than the \textit{Kepler} sample. This supports the observation that younger populations of $\delta$\,Sct stars have higher pulsator fractions.


\section{Discussion}

\subsection{Measuring $\Delta\nu$}
\label{ssec:dnu_method}

As noted by \citet{murphyetal2023}, $\delta$\,Sct stars of a given metallicity have very similar densities near the ZAMS, and hence similar values of $\Delta\nu$, despite their spread in mass. We therefore expect stars in an association like Cep--Her to have similar densities, although there are several effects that would produce a spread. The most obvious of these is rotation. Centrifugal deformation produces a marked reduction in stellar density, which has been recently demonstrated through $\Delta\nu$ measurements with K2 and TESS data of the Pleiades \citep{murphyetal2022,beddingetal2023}. 
Another factor relevant to the discussion is binarity. Stars can appear over-luminous on the H--R diagram because they have unresolved binary companions, and a star's $\Delta\nu$ can indicate that the true stellar density is higher than the H--R diagram position would suggest.

We again focus on the ZAMS group, since these stars are better characterised and can be assumed to members of the association. We also examined stars that lie above the ZAMS group, but found that very few of these stars had regular patterns that would permit a measurement of $\Delta\nu$. We attempted to measure $\Delta\nu$ for all 126 stars in the ZAMS group by constructing \'echelle diagrams following the method of \citet{beddingetal2020}, namely, adjusting $\Delta\nu$ until vertical ridges of modes are found in the \'echelles. During this process, and for this purpose only, we smoothed the Fourier data to a FHWM of 0.05\,d$^{-1}$, because different stars otherwise have different frequency resolutions depending on how much TESS data they have. This target resolution was chosen to equal the approximate $\Delta\nu$ precision attainable on a best-case basis ($\pm0.02$\,d$^{-1}$). Unlike the \citet{beddingetal2020} sample, the Cep--Her stars do not all have regular patterns (the ridges more often have gaps), hence we estimated our $\Delta\nu$ precision to be slightly lower, at $\pm0.03$\,d$^{-1}$, which is still adequate for our purposes. Of those 126 stars, 70 had measurable $\Delta\nu$ values, 34 had too few peaks to produce vertical ridges in the \'echelles, and 22 had many peaks but no discernible $\Delta\nu$ (Table\:\ref{tab:big}).

\subsection{Evaluating the reliability of Gaia {\tt vbroad}}
\label{ssec:vbroad}

In order to distinguish evolutionary effects from rotational ones, we need reliable measurements of the stellar rotation. For this, we used the Gaia {\tt vbroad} parameter, which is a measure of the total line broadening, hence is a projected quantity along the line of sight, like $v \sin i$. In this subsection, we evaluate whether {\tt vbroad} is a suitable substitute for $v \sin i$.

The Gaia Team has already characterised the {\tt vbroad} parameter available in DR3 in detail \citep{frematetal2023}. That study characterised {\tt vbroad} for a range of stellar temperatures and brightnesses and explored the parameter's technical details. Specifically, it examined behaviour either side of a temperature reference point of 7500\,K, which happens to be near the middle of our range of interest. Here, we focus just on the A/F stars to estimate the reliability of {\tt vbroad} for $\delta$\,Sct pulsators.

The {\tt vbroad} parameter is not a direct measurement of rotation, but rather all the line broadening present in the calcium triplet located within the 846--870\,nm spectral range of the RVS spectrograph \citep{cropperetal2018,frematetal2023}. As noted in Sec.\,\ref{sec:methods}, $\delta$\,Sct stars are typically rapid rotators so rotation is expected to be the dominant contributor, but the instrumental broadening ($R=11500$, $\sim$26\,km\,s$^{-1}$) should not be forgotten. Micro- and macroturbulence should be insignificant for most $\delta$\,Sct stars \citep{aertsetal2009,landstreetetal2009,doyleetal2014,grassitellietal2015b} and, in any case, no macroturbulence is accounted for in the template spectra used in the {\tt vbroad} calculation \citep{frematetal2023}. Especially for hot stars, the overall accuracy is sensitive to the quality of the match of those template spectra, and inaccuracies of 500\,K in the template spectrum can have large detrimental effects on the resulting {\tt vbroad} fit (see \citealt{frematetal2023} for further details).\footnote{We note, therefore, that $\nu_{\rm max}$-derived temperatures would be insufficiently accurate for applications such as this.}

We evaluated the Gaia {\tt vbroad} measurements against a benchmark sample of rotational velocities for B/A/F stars \citep{zorec&royer2012}, catalogued from measurements made by \citet{royeretal2007} wherein the details of their method for determining $v\sin i$ can be found. From this sample, we removed stars they had flagged as close binaries, so as not to bias results. We note that the Gaia team also removed {\tt vbroad} measurements for stars detected as SB2s \citep{katzetal2023}. After cross-matching the HD numbers from the \citet{zorec&royer2012} catalogue with Gaia IDs with {\tt astroquery.simbad}, we used {\tt astroquery.gaia} to obtain {\tt phot\_mean\_g\_mag}, $G_{\rm BP} - G_{\rm RP}$ colour, {\tt ruwe}, {\tt vbroad}, and {\tt vbroad\_error}. We removed from the sample any stars without {\tt vbroad} measurements. Since we are more interested in any systematic biases than in random errors, we also removed stars with fractional errors ({\tt vbroad\_error/vbroad}) greater than 10\%. We note that too few stars had Gaia {\tt vsini$_{\rm esphs}$} for a useful comparison to be drawn, so our focus remained on {\tt vbroad}. 

\begin{figure}
\begin{center}
\includegraphics[width=0.48\textwidth]{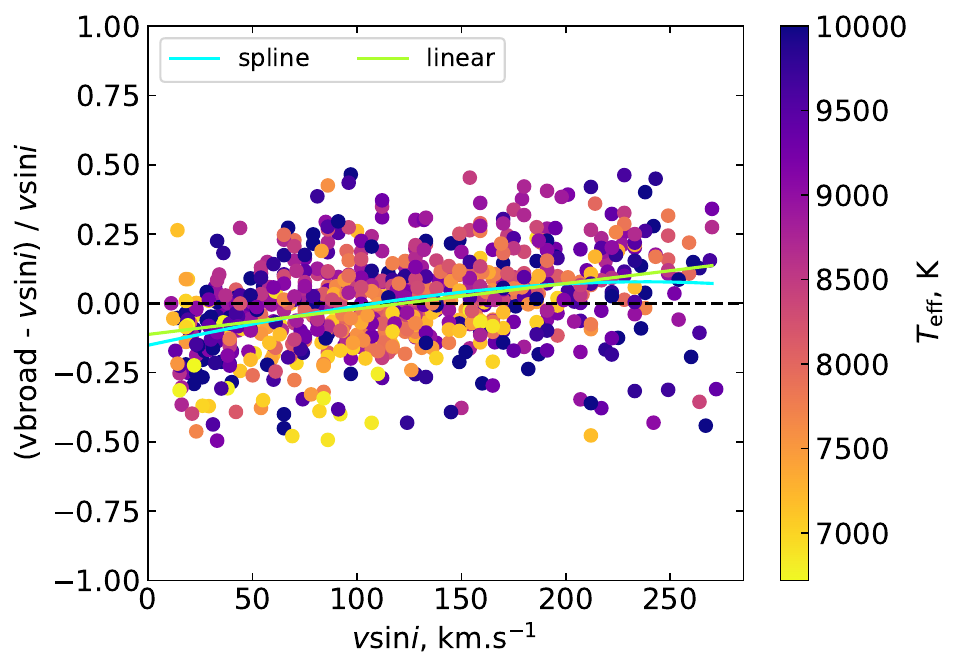}
\caption{Dependence of {\tt vbroad} accuracy on known rotation rates ($v\sin i$ from \citealt{zorec&royer2012}) and on effective temperature (from the TIC, \citealt{stassunetal2019}). The black and green solid lines are linear and univariate spline fits to the data, respectively.}
\label{fig:velocity}
\end{center}
\end{figure}

Fig.\,\ref{fig:velocity} shows that {\tt vbroad} tends to be 10--15\% smaller than $v \sin i$ when the star is a slow rotator ($\lesssim50$\,km\,s$^{-1}$), performs well between 75 and 200 \,km\,s$^{-1}$, but overestimates rotation velocities for rapid rotators ($v\sin i>200$\,km\,s$^{-1}$). It also shows no strong temperature dependence, but the scatter is larger for stars near or above 10,000\,K. The $\delta$\,Sct instability strip blue edge lies at around 9000\,K \citep{dupretetal2005b,xiongetal2016,murphyetal2019}, hence this should be unimportant for $\delta$\,Sct stars generally. In conclusion, {\tt vbroad} is a reliable indicator of rotation velocities for $\delta$\,Sct stars, but underestimates the rotation velocity of slow rotators ($v\sin i$ $\lesssim50$\,km\,s$^{-1}$) by 10--15\%.

\subsection{The rotation distribution of stars in the ZAMS group}

\begin{figure}
\begin{center}
\includegraphics[width=0.48\textwidth]{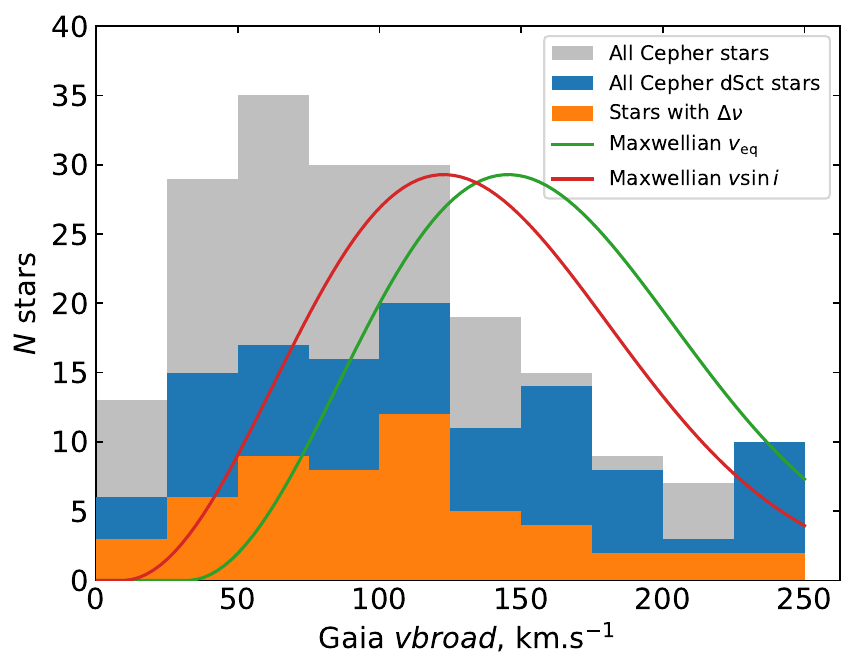}
\caption{Gaia {\tt vbroad} distribution for Cep--Her stars, based on pulsation properties. All three samples have been confined to the colour where $\delta$\,Sct stars are found. The grey histogram shows all Cep--Her stars in the ZAMS group, the blue histogram shows those that are $\delta$\,Sct stars, and the orange histogram shows $\delta$~Scuti stars for which a $\Delta\nu$ value could be measured. The green curve is the Maxwellian distribution of equatorial rotation velocities from \citet{zorec&royer2012} for stars of mass 1.6 to 2.0\,M$_{\odot}$. Note that close binaries and chemically peculiar stars are removed from that sample to give the distribution for (superficially) normal, single stars. The red curve is the same distribution divided by $\sqrt{3}/2$ to (crudely) reproject it into an observed ($v \sin i$) distribution.}
\label{fig:cepher_vbroad}
\end{center}
\end{figure}

Having established that {\tt vbroad} is sufficiently reliable, we consider the distribution of rotation velocities of stars in the ZAMS group of Cep--Her. We are interested in any dependence of pulsation properties on rotation velocity, so we defined three phenomenological groups: (i) the full ZAMS-group sample of Cep--Her stars; (ii) the subset of group (i) that pulsate; and (iii) the subset of group (ii) that have measurable $\Delta\nu$. To limit mass-dependent effects on the {\tt vbroad} distribution, we limit all three groups to the colour range where most pulsators are found, namely $0.08<(G_{\rm BP} - G_{\rm RP})_0<0.42$. We show their {\tt vbroad} distribution in Fig.\,\ref{fig:cepher_vbroad}.

It appears that more rapid rotators are more likely to pulsate as $\delta$~Sct stars, consistent with the recent finding by \citet{gootkinetal2024} who used an all-sky sample of TESS $\delta$~Scuti stars. We find it is possible to infer $\Delta\nu$ for approximately half of the $\delta$~Sct stars with {\tt vbroad} $<$ 150\,km\,s$^{-1}$. Above this velocity, $\Delta\nu$ is only apparent for around 1 in 4. Hence, while more rapid rotators are more likely to pulsate, there is less order (regularity) to their mode frequencies.

Overall, the observed {\tt vbroad} distribution of stars in CepHer is slower than expected from observations of field stars from \citet{zorec&royer2012}. An explanation for this is lacking -- the systematic underestimation of $v \sin i$ by the {\tt vbroad} parameter is only 10--15\% and vanishes by $v \sin i = 100$\,km\,s$^{-1}$ (Sec.\,\ref{ssec:vbroad}), hence some other explanation is required.

\subsection{The effect of rotation on pulsation}
\label{ssec:rot-puls}

\begin{figure}
\begin{center}
\includegraphics[width=0.48\textwidth]{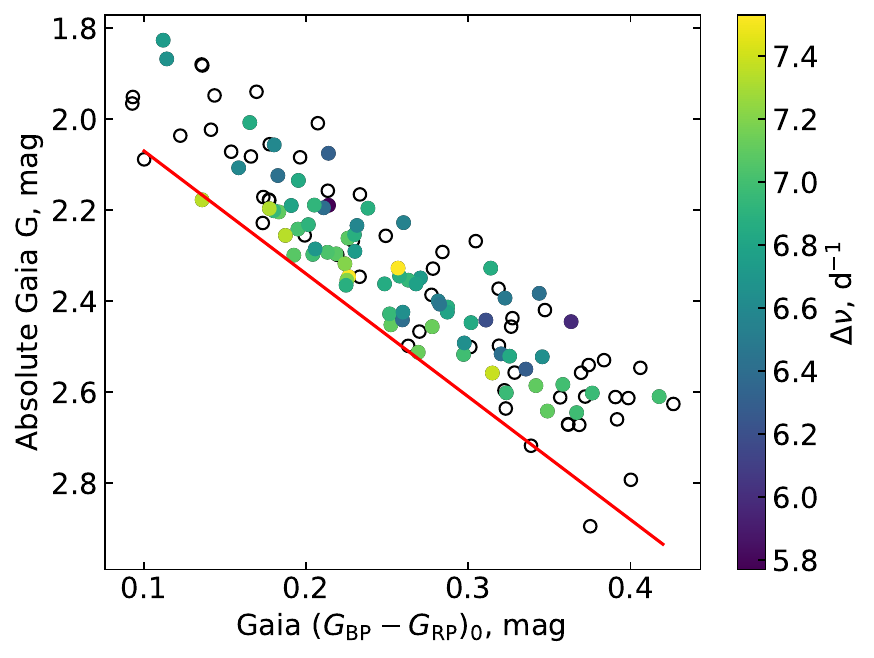}
\includegraphics[width=0.48\textwidth]{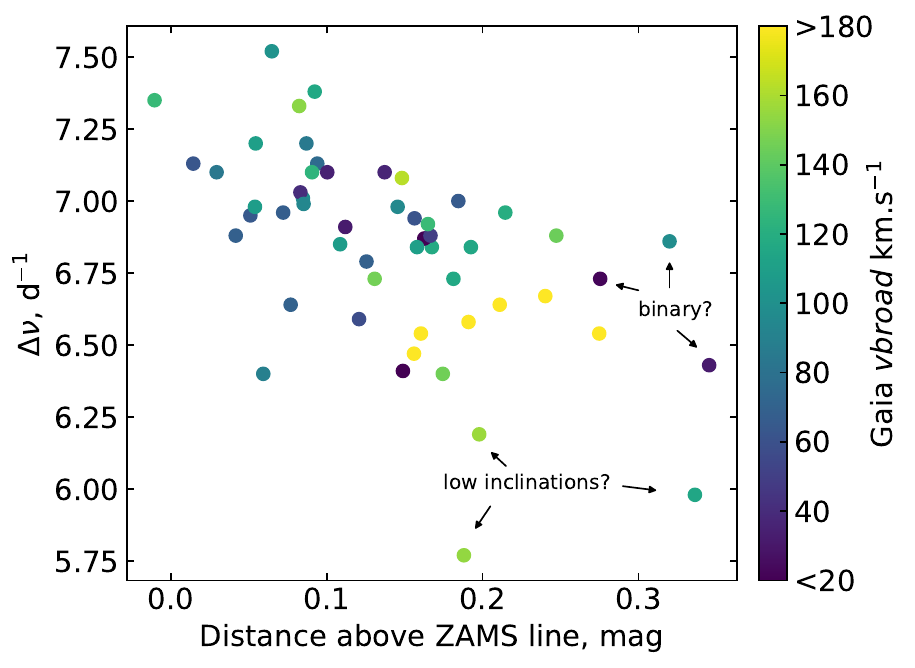}
\caption{Top: The colour--magnitude diagram for the ZAMS group of stars in Cep--Her, colour-coded by measured $\Delta\nu$. Open black symbols show stars for which no $\Delta\nu$ could be measured, 40\% of which are rapid rotators. The red line is the ZAMS line introduced in Sec.\,\ref{ssec:pack}. Bottom: the subset of stars for which $\Delta\nu$ was measurable and Gaia {\tt vbroad} was available, showing $\Delta\nu$ as a function of the height above the ZAMS line. Stars farther from the ZAMS line should have a lower density, either due to rapid rotation or due to evolutionary effects such as pre-MS status. Stars indicated with arrows are discussed in the text.}
\label{fig:dnu}
\end{center}
\end{figure}

The $\Delta\nu$ values measured in Sec.\,\ref{ssec:dnu_method} are tabulated in Table\:\ref{tab:big} and shown in Fig.\,\ref{fig:dnu} (top), wherein two trends are apparent. Firstly, stars with the lowest densities (lowest $\Delta\nu$) tend to lie farther above the ZAMS, as expected. We will attempt to separate the causes of those low densities shortly. Secondly, stars without a clear $\Delta\nu$ are preferentially those farther from the ZAMS line. We found that 40\% of those are rapid rotators ({\tt vbroad} in excess of 100\,km\,s$^{-1}$), confirming the oft-stated speculation that rapid rotation spoils the regular patterns (Sec.\,\ref{sec:scaling} and references therein).

Fig.\,\ref{fig:dnu} (bottom) shows that, as Gaia {\tt vbroad} increases, the distance above the ZAMS line generally increases and $\Delta\nu$ generally decreases. There are a few noteworthy exceptions. Three stars at the bottom of the diagram, having the lowest $\Delta\nu$, lie far above the ZAMS yet have only moderately high {\tt vbroad}. These stars are TIC\,17372709, TIC\,158216795, and TIC\,171884646. They are probably rapid rotators that are seen at low or moderate inclinations. There are also three stars on the right, lying particularly far above the ZAMS line, with moderate $\Delta\nu$ and low {\tt vbroad} (TIC\,27978717, TIC\,135412676, and TIC\,322497193 with v{\tt broad} = 22, 97, and 31\,km\,s$^{-1}$, respectively). They are unlikely to be rapid rotators seen at low inclination because they do not lie towards the bottom of the diagram (lower density). The obvious explanation is a binary companion that would cause these stars to appear brighter without being less dense. Hence, even with relatively simple asteroseismic data ($\Delta\nu$) we can infer when stars are rapid rotators seen pole on, and can potentially flag binaries that might be missed in other data (e.g. Gaia {\tt ruwe}).


\section{Summary}

We have studied the $\delta$\,Sct stars in the Cep--Her association  to compare its pulsator fraction to that of associations or clusters of different ages, to evaluate the $\nu_{\rm max}$ scaling relation on a statistically significant homogenous sample, and to study the interplay between pulsation and rotation. We collated stellar properties from Gaia and from the TESS Input Catalogue, and found that there is a subsample of Cep--Her stars that form a tight pack (`the ZAMS group') in a colour-magnitude diagram. That subsample should be free of large metallicity or age variations, and we analysed its stars in detail.

We employed four different methods to measure $\nu_{\rm max}$, thereby minimising any methodological (systematic) bias in the measurements. We found a correlation between dereddened Gaia $(G_{\rm BP} - G_{\rm RP})_0$ colour and $\nu_{\rm max}$, whether examined as the range of measured $\nu_{\rm max}$ values for each star, or as their mean. However, we found substantial scatter in that relation of 372\,K at $1\sigma$, even after an extensive outlier removal effort. This scatter is much larger ($>$1000\,K) if we do not exclude outliers or restrict our sample to the ZAMS group. Given that the instability strip is 2000\,K wide, we surmise that the $\nu_{\rm max}$ relation is of little practical use.

Nonetheless, by combining measurements from multiple clusters we observed {\it some} structure in the $\nu_{\rm max}$--$T_{\rm eff}$ diagram. Specifically, we found two ridges that bear similarity to those in the Period--luminosity diagram. A closer analysis of Gaia {\tt vbroad} values along the ridges suggests (though only at 1$\sigma$) that rapid rotation causes stars to pulsate in lower radial orders, explaining the temperature-independent ridge of the young stars comprising our sample.

The pulsator fraction in Cep--Her peaks at 100\%, at \mbox{$(G_{\rm BP} - G_{\rm RP})_0$ = 0.30--0.35\,mag,} corresponding to $7750 \gtrsim T_{\rm eff} \gtrsim 7500$\,K. This is the first such measurement for an association or cluster younger than 100\,Myr, and forms part of a continuing effort to better characterise the fraction of stars in the instability strip that pulsate. That effort began with \textit{Kepler} and has continued with clusters of different ages observed by TESS. The emerging trend is that the pulsator fraction is higher for younger $\delta$\,Sct stars, which we attribute to the onset of helium settling in young main-sequence A stars even before 100\,Myr.

We evaluated the Gaia {\tt vbroad} parameter against  $v\sin i$ for an independent sample of A stars and concluded that {\tt vbroad} is a reliable estimator of rotation velocities for $\delta$\,Sct stars generally, but underestimates the rotation velocities of slow rotators ($v\sin i < 50$\,km\,s$^{-1}$) by 10--15\%. We found that more rapid rotators in the instability strip are more likely to pulsate as $\delta$\,Sct stars, albeit with less regular pulsation patterns, and that Cep--Her contains an unexplained excess of slow rotators.

We were able to measure the asteroseismic large spacing, $\Delta\nu$, for 70 of the 126 stars in the ZAMS group via the \'echelle method. 
We observed that the pulsators for which $\Delta\nu$ could not be measured were preferentially the stars farthest from the ZAMS (i.e. stars of the lowest density). 
At least 40\% of these were rapid rotators (high {\tt vbroad}), hence our conclusion that rapid rotation does indeed spoil the regular patterns of modes in $\delta$\,Sct stars. 
We also showed that a correlation between $\Delta\nu$ and rotation (via the proxy quantity `distance above the ZAMS' in magnitudes), allows rapid rotators seen at low inclinations to be distinguished from slow rotators, and may assist with the identification of unresolved binaries.

Future work will include mode identification and the application of rotating models to determine asteroseismic ages for different subgroups in the Cep--Her Complex. Spectroscopic observations are ongoing, which will offer a narrow metallicity prior to hone the results. We expect to apply hierarchical Bayesian modelling to a dozen or so members of the Complex in this manner, refining both relative and absolute ages of the subgroups of Cep--Her.


\section*{Acknowledgements}

We thank Daniel Foreman-Mackey for expanding on his {\tt numpyro} tutorial and helping to apply it to the $\nu_{\rm max}$ data analysed in Fig.\,\ref{fig:lines}.
SJM was supported by the Australian Research Council (ARC) through Future Fellowship FT210100485. TRB was also supported by the ARC, through DP210103119 and FL220100117.

\section*{Software}
We used the following software:
\begin{itemize}
    \item {\tt numpyro} \citep{numpyro1,numpyro2}
\item {\tt jax} \citep{jax1}
\item {\tt arviz} \citep{arviz1}
\item {\tt lightkurve} \citep{lightkurvecollaboration2018}
\item {\tt astroquery} \citep{astroquery1}
\end{itemize}

\section*{Data Availability}

\noindent
We make our stellar parameters table (Table\:\ref{tab:big}) available online as a csv file accompanying this article. The table of the five strongest pulsation peaks (Table\:\ref{tab:modes}) is similarly available as a csv, and we provide a separate repository of all peaks (not just the first five) for each star. TESS data are publicly available online via the Mikulski Archive for Space Telescopes (MAST), and are readily accessible via the {\tt lightkurve} package. 


\bibliographystyle{mnras}
\interlinepenalty=10000
\bibliography{sjm_bibliography} 



\appendix
\section{Data Tables}
\label{app}

\begin{landscape}

\begin{table}
\caption{Data for stars in the ZAMS group. The columns are as follows: (i) TIC number, (ii) HD number, where available, (iii) the TESS cadence used, (iv) the Gaia mean apparent $g$ magnitude, (v) the Gaia absolute G magnitude, (vi) the dereddened Gaia $(G_{\rm BP} - G_{\rm RP})_0$ colour, (vii) the vertical distance above the ZAMS line (see Sec.\,\ref{ssec:pack}), (viii) the effective temperature from the TIC and (ix) its uncertainty, (x) the Gaia renormalised unit weight error ({\tt ruwe}), (xi) the Gaia {\tt vbroad} parameter and (xii) its uncertainty {\tt vbroad\_error} from the Gaia source catalogue, (xiii) the frequency of highest amplitude and (xiv) its corresponding amplitude, (xv) the $\nu_{\rm max}$ value via the moment method using amplitude and (xvi) using power, (xvii) the $\nu_{\rm max}$ value via the smoothing method using amplitude and (xviii) using power, and finally (xix) $\Delta\nu$ ($\pm 0.03$\,d$^{-1}$), which was not measurable for all stars (refer to Sec.\,\ref{ssec:dnu_method}).}
\label{tab:big}
\begin{tabular}{rrrrrrrrrrrrrrrrrrr}
\toprule
\multicolumn{14}{c}{} & \multicolumn{2}{c}{\underline{$\nu_{\rm max}$ (moment)}} & \multicolumn{2}{c}{\underline{$\nu_{\rm max}$ (smooth)}} & \\
TIC & HD & cadence & $g$ & $G$ & $(G_{\rm BP} - G_{\rm RP})_0$ & {\tt vert} & $T_{\rm eff}$ & e\_ $T_{\rm eff}$ & {\tt ruwe} & {\tt vbroad} & e\_{\tt vbroad} & $f_{\rm max}$ & A$_{\rm max}$ & amp. & pow. & amp. & pow. & $\Delta\nu$ \\
 &  & min & mag & mag & mag & mag & K & K & & km\,s$^{-1}$ & km\,s$^{-1}$ & d$^{-1}$ & ppt & d$^{-1}$& d$^{-1}$ & d$^{-1}$ & d$^{-1}$ & d$^{-1}$ \\
\midrule
7992735 & 348886 & 10 & 10.266 & 2.61 & 0.40 & 0.26 & 7163 & 174 & 0.99 & 26.04 & 4.35 & 19.27 & 1.091 & 18.91 & 19.37 & 18.24 & 19.43 & -- \\
8598196 & 349008 & 10 & 10.502 & 2.61 & 0.36 & 0.15 & 7390 & 191 & 1.32 & 47.30 & 4.14 & 21.112 & 0.544 & 24.00 & 22.48 & 23.23 & 22.82 & -- \\
11218613 & 228591 & 2 & 10.216 & 2.35 & 0.23 & 0.06 & 8085 & 188 & 1.01 & 102.08 & 20.25 & 35.672 & 1.275 & 49.21 & 46.49 & 50.49 & 45.81 & 7.52 \\
11588737 & 228700 & 10 & 9.690 & 1.88 & 0.14 & 0.29 & 8673 & 206 & 1.29 & 194.54 & 14.39 & 32.819 & 4.574 & 26.20 & 31.52 & 28.53 & 32.03 & -- \\
11708996 & 228681 & 2 & 9.926 & 2.18 & 0.18 & 0.10 & 8356 & 123 & 0.82 & 154.88 & 12.89 & 44.398 & 2.144 & 38.41 & 45.08 & 45.40 & 46.31 & -- \\
12361752 & 228838 & 2 & 10.068 & 2.29 & 0.23 & 0.13 & 8031 & 155 & 0.87 & 145.81 & 14.02 & 44.871 & 1.216 & 44.15 & 44.82 & 44.58 & 44.86 & 6.73 \\
13876477 & 229189 & 10 & 10.016 & 2.16 & 0.21 & 0.22 & 8038 & 139 & 0.92 & 282.78 & 19.20 & 22.112 & 1.318 & 23.26 & 22.09 & 23.31 & 21.57 & -- \\
15247229 & -- & 2 & 9.997 & 2.18 & 0.18 & 0.10 & 8263 & 157 & 1.10 & 105.01 & 8.18 & 49.984 & 1.272 & 49.71 & 49.63 & 49.78 & 49.74 & -- \\
15478761 & -- & 10 & 10.501 & 2.67 & 0.36 & 0.11 & 7397 & 158 & 1.10 & 22.94 & 5.91 & 19.29 & 0.126 & 16.83 & 17.18 & 16.28 & 21.67 & -- \\
16310676 & -- & 10 & 9.633 & 1.95 & 0.09 & 0.10 & 8683 & 123 & 1.92 & 39.06 & 3.77 & 43.399 & 0.248 & 38.47 & 41.59 & 38.85 & 43.07 & -- \\
16621585 & -- & 2 & 10.455 & 2.64 & 0.32 & 0.04 & 7552 & 147 & 0.94 & 46.25 & 9.40 & 27.964 & 0.775 & 25.30 & 25.55 & 24.93 & 25.58 & -- \\
16622862 & -- & 2 & 10.494 & 2.72 & 0.34 & 0.00 & 7568 & 136 & 1.14 & 45.10 & 5.41 & 23.185 & 0.685 & 17.26 & 14.72 & 16.15 & 14.02 & -- \\
16739247 & -- & 2 & 10.358 & 2.56 & 0.31 & 0.09 & 7706 & 178 & 0.99 & 115.74 & 14.11 & 22.909 & 1.563 & 28.79 & 26.93 & 28.16 & 26.95 & 7.38 \\
17372709 & -- & 2 & 10.040 & 2.19 & 0.21 & 0.19 & 8189 & 198 & 0.90 & 154.14 & 9.60 & 18.286 & 0.903 & 28.55 & 25.94 & 27.46 & 25.44 & 5.77 \\
20774645 & -- & 10 & 10.217 & 2.56 & 0.37 & 0.24 & 7356 & 134 & 1.01 & -- & -- & 28.404 & 0.11 & 24.17 & 25.30 & 10.00 & 34.92 & -- \\
20819550 & -- & 2 & 10.508 & 2.45 & 0.25 & 0.03 & 7865 & 131 & 1.12 & 83.78 & 11.09 & 41.615 & 0.764 & 44.20 & 42.66 & 43.01 & 42.15 & 7.10 \\
21150042 & -- & 2 & 10.679 & 2.60 & 0.32 & 0.07 & 7584 & 143 & 0.87 & 68.31 & 8.42 & 22.508 & 1.061 & 27.84 & 25.12 & 26.92 & 25.36 & -- \\
27978717 & -- & 10 & 9.846 & 1.83 & 0.11 & 0.28 & 8545 & 136 & 1.16 & 22.30 & 4.31 & 39.521 & 0.46 & 33.69 & 35.68 & 36.43 & 38.06 & 6.73 \\
28397009 & -- & 2 & 10.082 & 2.59 & 0.34 & 0.14 & 7490 & 126 & 1.23 & 36.34 & 3.45 & 21.595 & 1.129 & 28.60 & 25.70 & 26.96 & 24.38 & 7.10 \\
28400691 & -- & 10 & 10.355 & 2.53 & 0.38 & 0.31 & 7338 & 135 & 6.03 & 85.81 & 10.68 & 21.825 & 0.359 & 21.50 & 21.67 & 21.10 & 22.88 & -- \\
28679106 & -- & 2 & 9.708 & 2.35 & 0.23 & 0.05 & 8094 & 145 & 1.06 & 109.90 & 7.08 & 50.842 & 1.584 & 52.57 & 53.32 & 53.06 & 53.70 & 7.20 \\
29539591 & -- & 10 & 9.896 & 2.01 & 0.21 & 0.35 & 8231 & 169 & 2.41 & 155.85 & 19.72 & 52.032 & 0.464 & 53.88 & 53.43 & 54.27 & 52.96 & -- \\
40485583 & 227470 & 2 & 9.770 & 1.95 & 0.14 & 0.24 & 8538 & 177 & 0.94 & -- & -- & 62.558 & 0.075 & 63.21 & 62.34 & 62.02 & 57.27 & -- \\
43795649 & -- & 2 & 9.726 & 2.30 & 0.22 & 0.09 & 7839 & 129 & 0.98 & -- & -- & 38.535 & 1.094 & 33.90 & 34.90 & 35.07 & 35.61 & 7.07 \\
63819901 & -- & 2 & 10.183 & 2.50 & 0.26 & 0.01 & 8508 & 160 & 1.05 & 107.99 & 14.15 & 21.154 & 1.593 & 33.03 & 29.38 & 31.34 & 26.70 & -- \\
64050903 & -- & 2 & 10.161 & 2.43 & 0.25 & 0.05 & 8599 & 351 & 1.04 & 64.04 & 5.52 & 50.16 & 0.42 & 41.59 & 43.28 & 15.82 & 14.12 & 6.95 \\
66585916 & -- & 2 & 9.945 & 2.24 & 0.19 & 0.08 & 8753 & 326 & 0.98 & 98.66 & 7.98 & 52.829 & 1.047 & 50.00 & 50.90 & 50.39 & 51.24 & 7.01 \\
68757401 & -- & 10 & 10.278 & 2.61 & 0.39 & 0.24 & 7199 & 129 & 3.54 & 238.96 & 15.02 & 21.767 & 0.126 & 24.43 & 24.16 & 22.85 & 28.88 & -- \\
68766561 & -- & 2 & 9.598 & 2.26 & 0.20 & 0.08 & 8125 & 135 & 1.95 & 131.72 & 10.28 & 53.134 & 1.497 & 49.97 & 51.73 & 51.39 & 52.66 & -- \\
91212494 & 227853 & 2 & 10.069 & 2.45 & 0.30 & 0.17 & 7662 & 141 & 0.94 & 117.32 & 6.01 & 34.528 & 0.799 & 30.16 & 30.68 & 30.40 & 31.74 & 6.84 \\
91825479 & 228034 & 2 & 9.702 & 2.07 & 0.21 & 0.30 & 8887 & 464 & 1.79 & -- & -- & 43.884 & 1.181 & 44.00 & 43.86 & 44.18 & 44.05 & 6.31 \\
92428258 & 228216 & 2 & 9.613 & 2.01 & 0.17 & 0.24 & 8653 & 361 & 0.97 & -- & -- & 38.679 & 1.705 & 33.45 & 34.23 & 33.23 & 35.12 & 6.85 \\
111560278 & -- & 10 & 10.354 & 2.66 & 0.39 & 0.20 & 7271 & 332 & 0.95 & 57.35 & 6.24 & 23.301 & 0.738 & 28.22 & 28.95 & 31.65 & 29.87 & -- \\
120257786 & -- & 2 & 10.090 & 2.35 & 0.27 & 0.18 & 7899 & 141 & 1.39 & 115.64 & 9.52 & 31.584 & 0.762 & 41.67 & 40.03 & 41.05 & 39.91 & 6.73 \\
120361047 & -- & 2 & 10.358 & 2.52 & 0.35 & 0.21 & 7469 & 122 & 1.06 & 240.67 & 9.07 & 29.878 & 0.593 & 28.65 & 27.14 & 27.31 & 27.15 & 6.64 \\
120824631 & -- & 10 & 10.295 & 2.64 & 0.37 & 0.15 & 7500 & 137 & 0.98 & 95.51 & 8.66 & 24.499 & 0.473 & 22.88 & 23.46 & 23.45 & 24.22 & 6.98 \\
120827848 & -- & 2 & 9.889 & 2.12 & 0.18 & 0.17 & 8274 & 151 & 1.00 & -- & -- & 48.041 & 1.585 & 46.20 & 47.48 & 47.20 & 47.64 & 6.42 \\
121275246 & -- & 2 & 10.193 & 2.44 & 0.26 & 0.06 & 8036 & 147 & 0.93 & 88.94 & 8.71 & 33.269 & 1.009 & 39.30 & 36.85 & 38.38 & 36.25 & 6.40 \\
121537103 & -- & 2 & 10.004 & 2.27 & 0.23 & 0.15 & 8106 & 136 & 0.84 & 105.50 & 6.01 & 48.192 & 0.821 & 44.84 & 45.92 & 45.83 & 46.74 & -- \\
121869336 & -- & 2 & 10.316 & 2.35 & 0.26 & 0.16 & 7963 & 152 & 1.01 & 61.51 & 6.33 & 45.211 & 1.562 & 34.90 & 39.14 & 36.36 & 40.39 & 6.94 \\

\end{tabular}
\end{table}
\end{landscape}

\begin{landscape}
\begin{table}
\contcaption{}
\begin{tabular}{rrrrrrrrrrrrrrrrrrr}
\toprule
\multicolumn{14}{c}{} & \multicolumn{2}{c}{\underline{$\nu_{\rm max}$ (moment)}} & \multicolumn{2}{c}{\underline{$\nu_{\rm max}$ (smooth)}} & \\
TIC & HD & cadence & $g$ & $G$ & $(G_{\rm BP} - G_{\rm RP})_0$ & {\tt vert} & $T_{\rm eff}$ & e\_ $T_{\rm eff}$ & {\tt ruwe} & {\tt vbroad} & e\_{\tt vbroad} & $f_{\rm max}$ & A$_{\rm max}$ & amp. & pow. & amp. & pow. & $\Delta\nu$ \\
 &  & min & mag & mag & mag & mag & K & K & & km\,s$^{-1}$ & km\,s$^{-1}$ & d$^{-1}$ & ppt & d$^{-1}$& d$^{-1}$ & d$^{-1}$ & d$^{-1}$ & d$^{-1}$ \\
\midrule
122132758 & -- & 2 & 10.124 & 2.27 & 0.30 & 0.35 & 7701 & 127 & 4.56 & 165.31 & 11.47 & 22.102 & 0.739 & 28.90 & 28.51 & 28.16 & 27.38 & -- \\
122673431 & -- & 2 & 10.294 & 2.41 & 0.29 & 0.16 & 7849 & 155 & 1.03 & 18.27 & 2.93 & 41.574 & 2.182 & 29.24 & 32.81 & 28.33 & 37.39 & 6.87 \\
122715830 & -- & 2 & 9.651 & 1.87 & 0.11 & 0.24 & 8833 & 138 & 0.76 & 192.10 & 13.63 & 46.76 & 0.86 & 44.06 & 43.88 & 44.45 & 44.51 & 6.67 \\
135412676 & 347856 & 2 & 10.055 & 2.33 & 0.31 & 0.32 & 7507 & 147 & 3.87 & 97.62 & 10.60 & 40.727 & 0.342 & 38.12 & 38.21 & 38.68 & 38.68 & 6.86 \\
136708821 & -- & 2 & 9.969 & 2.39 & 0.32 & 0.28 & 7650 & 283 & 2.34 & -- & -- & 17.54 & 0.819 & 29.01 & 26.78 & 28.15 & 25.16 & 6.48 \\
137214324 & -- & 2 & 10.484 & 2.67 & 0.36 & 0.11 & 7363 & 122 & 0.99 & 46.05 & 4.17 & 22.809 & 1.442 & 18.97 & 20.31 & 18.69 & 20.58 & -- \\
137554985 & -- & 10 & 10.106 & 2.55 & 0.41 & 0.35 & 7231 & 124 & 0.92 & 76.71 & 5.51 & 18.325 & 0.898 & 16.20 & 17.66 & 16.31 & 17.64 & -- \\
157202091 & -- & 2 & 9.708 & 2.29 & 0.21 & 0.08 & 7961 & 150 & 0.97 & 40.58 & 4.95 & 42.148 & 1.46 & 46.50 & 46.07 & 46.16 & 45.95 & 7.03 \\
158082591 & 348246 & 2 & 9.827 & 2.33 & 0.28 & 0.22 & 7776 & 190 & 1.31 & 244.32 & 21.73 & 24.128 & 0.929 & 28.00 & 26.01 & 26.67 & 25.11 & -- \\
158082593 & 348247 & 2 & 9.613 & 2.11 & 0.16 & 0.12 & 8381 & 209 & 0.89 & 57.32 & 4.59 & 24.128 & 0.341 & 26.54 & 24.95 & 25.06 & 25.17 & 6.59 \\
158186602 & 348233 & 10 & 10.680 & 2.63 & 0.43 & 0.33 & 7088 & 150 & 1.14 & 27.11 & 8.77 & 26.183 & 0.217 & 22.89 & 24.19 & 17.45 & 26.43 & -- \\
158216795 & -- & 2 & 10.215 & 2.44 & 0.36 & 0.34 & 7200 & 135 & 0.92 & 114.70 & 8.40 & 17.827 & 1.60 & 22.78 & 22.59 & 21.52 & 22.11 & 5.98 \\
159719806 & 182952 & 2 & 9.630 & 2.09 & 0.10 & -0.02 & 8900 & 194 & 0.90 & 182.58 & 25.87 & 71.999 & 0.088 & 63.33 & 66.13 & 66.36 & 66.48 & -- \\
169461359 & 226029 & 10 & 10.376 & 2.61 & 0.42 & 0.32 & 7282 & 145 & 1.16 & -- & -- & 21.112 & 0.767 & 25.35 & 25.30 & 23.32 & 25.19 & 7.00 \\
169464993 & 226016 & 2 & 10.130 & 2.26 & 0.19 & 0.05 & 8315 & 179 & 1.00 & -- & -- & 55.205 & 4.802 & 52.29 & 57.15 & 56.04 & 57.55 & 7.34 \\
170734156 & 226443 & 2 & 10.096 & 2.42 & 0.26 & 0.08 & 7858 & 138 & 0.92 & 69.98 & 12.71 & 34.408 & 1.185 & 40.85 & 38.11 & 39.79 & 37.66 & 6.64 \\
171591531 & 226631 & 2 & 10.460 & 2.46 & 0.28 & 0.09 & 7950 & 185 & 0.93 & 75.33 & 9.59 & 36.617 & 1.213 & 44.21 & 41.61 & 42.86 & 40.55 & 7.13 \\
171884646 & 226696 & 2 & 10.216 & 2.44 & 0.31 & 0.20 & 7783 & 210 & 0.85 & 156.11 & 18.29 & 34.892 & 0.825 & 31.78 & 28.83 & 31.81 & 29.03 & 6.19 \\
185495536 & -- & 2 & 9.721 & 2.39 & 0.28 & 0.16 & 8079 & 282 & 1.12 & -- & -- & 37.95 & 0.357 & 38.40 & 37.24 & 38.03 & 37.12 & -- \\
185953037 & 192119 & 2 & 9.950 & 2.20 & 0.18 & 0.09 & 8350 & 156 & 1.02 & 124.23 & 14.11 & 52.643 & 5.819 & 43.75 & 48.11 & 47.19 & 49.51 & 7.10 \\
187942715 & -- & 2 & 10.164 & 2.36 & 0.22 & 0.04 & 7746 & 202 & 1.06 & 68.66 & 4.72 & 20.967 & 2.607 & 24.84 & 22.79 & 23.62 & 21.96 & 6.88 \\
188058423 & -- & 2 & 10.376 & 2.52 & 0.32 & 0.15 & 7704 & 122 & 0.83 & 20.20 & 5.15 & 42.386 & 0.702 & 36.91 & 39.27 & 36.04 & 39.67 & 6.41 \\
188062922 & -- & 2 & 9.935 & 2.26 & 0.23 & 0.15 & 8057 & 123 & 0.99 & 163.00 & 14.86 & 36.658 & 1.022 & 35.40 & 34.67 & 34.77 & 35.01 & 7.08 \\
188892141 & -- & 2 & 10.270 & 2.19 & 0.21 & 0.17 & 8417 & 122 & 0.95 & 145.28 & 13.99 & 48.127 & 1.119 & 46.29 & 47.20 & 46.61 & 47.33 & 6.40 \\
189283102 & -- & 2 & 10.146 & 2.30 & 0.20 & 0.05 & 8110 & 122 & 0.98 & 108.41 & 12.27 & 51.707 & 1.177 & 49.43 & 49.82 & 50.06 & 50.10 & 6.98 \\
193074256 & -- & 10 & 9.984 & 2.58 & 0.36 & 0.18 & 7394 & 177 & 0.96 & 65.19 & 6.64 & 20.228 & 0.581 & 24.43 & 22.67 & 22.16 & 21.98 & 7.00 \\
193079746 & -- & 2 & 10.272 & 2.50 & 0.30 & 0.11 & 7682 & 140 & 0.98 & 146.78 & 13.35 & 27.601 & 0.48 & 26.00 & 25.99 & 25.80 & 26.20 & -- \\
194047659 & -- & 10 & 10.018 & 2.54 & 0.37 & 0.27 & 7685 & 216 & 1.36 & 173.55 & 18.33 & 27.133 & 0.387 & 26.35 & 25.78 & 25.98 & 25.67 & -- \\
195800821 & -- & 10 & 10.400 & 2.42 & 0.35 & 0.32 & 7756 & 340 & 2.35 & 117.16 & 20.81 & 34.972 & 0.199 & 33.32 & 33.96 & 46.61 & 38.64 & -- \\
198681209 & -- & 2 & 10.247 & 2.25 & 0.23 & 0.17 & 7981 & 232 & 0.86 & 49.13 & 15.43 & 46.077 & 1.998 & 44.11 & 45.82 & 45.04 & 45.90 & 6.88 \\
213150627 & -- & 2 & 9.892 & 2.02 & 0.14 & 0.16 & 8381 & 182 & 1.07 & 150.33 & 14.59 & 52.609 & 0.569 & 53.91 & 52.40 & 53.46 & 52.22 & -- \\
230446029 & -- & 2 & 9.882 & 2.36 & 0.25 & 0.11 & 7868 & 131 & 1.00 & 107.86 & 18.07 & 25.915 & 0.491 & 39.97 & 37.65 & 41.76 & 38.44 & 6.85 \\
235612665 & 348025 & 2 & 9.932 & 2.23 & 0.20 & 0.11 & 8024 & 129 & 0.88 & 31.53 & 3.05 & 52.662 & 1.282 & 44.83 & 48.26 & 46.87 & 48.98 & 6.91 \\
235722686 & 336104 & 10 & 10.463 & 2.64 & 0.35 & 0.10 & 7470 & 139 & 1.83 & 34.44 & 10.99 & 29.741 & 0.798 & 31.31 & 29.17 & 29.91 & 28.36 & 7.10 \\
239264213 & -- & 2 & 10.032 & 2.55 & 0.34 & 0.16 & 7521 & 144 & 1.14 & -- & -- & 19.548 & 0.896 & 26.89 & 25.05 & 26.66 & 24.56 & 6.29 \\
267764363 & 235364 & 2 & 9.680 & 1.97 & 0.09 & 0.08 & 9567 & 627 & 1.02 & -- & -- & 60.739 & 0.096 & 60.74 & 60.74 & 74.49 & 62.64 & -- \\
267859063 & 239506 & 2 & 9.887 & 2.30 & 0.22 & 0.09 & 8374 & 580 & 0.89 & 161.18 & 8.98 & 17.216 & 0.831 & 28.80 & 25.09 & 27.66 & 23.90 & -- \\
268490151 & -- & 2 & 9.995 & 2.08 & 0.20 & 0.25 & 8196 & 155 & 0.89 & -- & -- & 49.974 & 1.468 & 45.42 & 47.10 & 46.47 & 47.49 & -- \\
272708054 & -- & 2 & 9.660 & 2.17 & 0.23 & 0.26 & 8200 & 168 & 0.98 & 227.20 & 15.24 & 25.422 & 1.531 & 27.85 & 27.51 & 27.31 & 27.41 & -- \\
273580799 & -- & 10 & 10.299 & 2.79 & 0.40 & 0.09 & 7406 & 101 & 1.45 & 66.13 & 6.17 & 20.891 & 1.641 & 22.25 & 21.39 & 22.34 & 21.39 & -- \\
275432316 & -- & 2 & 10.105 & 2.60 & 0.32 & 0.07 & 7438 & 136 & 1.02 & 67.43 & 6.85 & 22.157 & 0.863 & 29.58 & 28.35 & 28.79 & 27.20 & 6.96 \\
276391310 & 239439 & 2 & 9.600 & 2.23 & 0.23 & 0.19 & 8139 & 219 & 1.01 & 234.17 & 14.66 & 18.418 & 1.143 & 26.07 & 20.69 & 23.37 & 19.14 & 6.58 \\
277096454 & 239470 & 2 & 9.700 & 2.35 & 0.23 & 0.08 & 8087 & 277 & 0.97 & 71.17 & 6.26 & 43.222 & 0.644 & 44.06 & 43.65 & 44.31 & 43.66 & -- \\
290222306 & -- & 2 & 9.961 & 2.29 & 0.21 & 0.07 & 8761 & 639 & 0.91 & -- & -- & 57.727 & 1.311 & 52.60 & 54.62 & 53.10 & 54.91 & 6.70 \\
 
\end{tabular}
\end{table}
\end{landscape}

\pagebreak

\begin{landscape}
\begin{table}
\contcaption{}
\begin{tabular}{rrrrrrrrrrrrrrrrrrr}
\toprule
\multicolumn{14}{c}{} & \multicolumn{2}{c}{\underline{$\nu_{\rm max}$ (moment)}} & \multicolumn{2}{c}{\underline{$\nu_{\rm max}$ (smooth)}} & \\
TIC & HD & cadence & $g$ & $G$ & $(G_{\rm BP} - G_{\rm RP})_0$ & {\tt vert} & $T_{\rm eff}$ & e\_ $T_{\rm eff}$ & {\tt ruwe} & {\tt vbroad} & e\_{\tt vbroad} & $f_{\rm max}$ & A$_{\rm max}$ & amp. & pow. & amp. & pow. & $\Delta\nu$ \\
 &  & min & mag & mag & mag & mag & K & K & & km\,s$^{-1}$ & km\,s$^{-1}$ & d$^{-1}$ & ppt & d$^{-1}$& d$^{-1}$ & d$^{-1}$ & d$^{-1}$ & d$^{-1}$ \\
\midrule
295019890 & -- & 2 & 10.528 & 2.51 & 0.27 & 0.01 & 8047 & 138 & 1.01 & 62.99 & 6.11 & 23.58 & 2.099 & 34.46 & 30.54 & 34.31 & 30.21 & 7.13 \\
295889569 & 235154 & 2 & 10.162 & 2.52 & 0.30 & 0.09 & 7778 & 156 & 1.05 & 94.35 & 10.03 & 22.266 & 1.265 & 27.05 & 25.86 & 26.84 & 25.48 & 6.99 \\
295893690 & -- & 2 & 9.876 & 2.34 & 0.26 & 0.15 & 7986 & 182 & 0.98 & -- & -- & 38.502 & 1.452 & 31.68 & 34.02 & 32.62 & 35.01 & 6.87 \\
297399778 & -- & 2 & 9.917 & 2.04 & 0.12 & 0.09 & 8805 & 303 & 0.95 & -- & -- & 53.457 & 0.25 & 54.69 & 53.88 & 58.36 & 54.14 & -- \\
303248599 & -- & 2 & 10.338 & 2.89 & 0.38 & -0.08 & 7459 & 161 & 3.25 & 23.81 & 5.63 & 40.66 & 0.518 & 41.57 & 41.34 & 40.36 & 41.27 & -- \\
310828006 & -- & 2 & 10.390 & 2.42 & 0.29 & 0.15 & 8173 & 291 & 0.93 & -- & -- & 20.732 & 1.949 & 27.48 & 24.07 & 26.07 & 23.00 & 6.79 \\
310832102 & -- & 2 & 10.246 & 2.32 & 0.22 & 0.09 & 8625 & 315 & 1.15 & 84.40 & 17.63 & 50.605 & 2.053 & 47.94 & 50.84 & 49.81 & 51.11 & 7.20 \\
313893742 & 335735 & 2 & 9.362 & 2.07 & 0.15 & 0.14 & 8325 & 152 & 0.93 & 78.21 & 6.30 & 59.398 & 0.926 & 55.57 & 56.32 & 55.89 & 56.33 & -- \\
317575159 & 336864 & 10 & 10.497 & 2.46 & 0.33 & 0.23 & 7505 & 138 & 1.16 & 247.66 & 19.26 & 18.31 & 0.619 & 22.83 & 20.30 & 19.33 & 19.96 & -- \\
320960331 & -- & 2 & 10.033 & 2.29 & 0.28 & 0.28 & 7715 & 136 & 0.92 & 106.38 & 6.22 & 26.106 & 1.305 & 22.98 & 24.84 & 24.00 & 25.15 & -- \\
321330636 & 239522 & 2 & 10.105 & 2.47 & 0.27 & 0.06 & 7988 & 480 & 1.14 & 106.54 & 7.72 & 21.45 & 2.386 & 28.90 & 23.19 & 25.63 & 22.44 & -- \\
322425940 & 336447 & 2 & 10.217 & 2.08 & 0.17 & 0.17 & 8323 & 153 & 1.08 & 201.69 & 18.26 & 51.011 & 1.253 & 47.55 & 47.13 & 46.89 & 47.62 & -- \\
322497193 & -- & 10 & 10.175 & 2.38 & 0.34 & 0.35 & 7887 & 255 & 0.87 & 31.45 & 3.90 & 42.214 & 1.232 & 32.12 & 34.87 & 32.72 & 35.91 & 6.43 \\
322708594 & 342552 & 10 & 10.192 & 2.52 & 0.33 & 0.16 & 7376 & 139 & 1.05 & 109.81 & 7.71 & 35.297 & 0.677 & 30.51 & 32.69 & 31.28 & 33.21 & 6.84 \\
332694657 & 335724 & 2 & 9.833 & 2.37 & 0.32 & 0.29 & 7721 & 171 & 1.04 & 49.47 & 4.72 & 21.333 & 0.493 & 25.80 & 25.66 & 25.52 & 25.64 & -- \\
332778908 & 166436 & 2 & 9.273 & 1.94 & 0.17 & 0.32 & 8291 & 142 & 1.07 & 180.75 & 14.68 & 32.09 & 2.105 & 25.57 & 29.12 & 27.26 & 30.22 & -- \\
333441069 & 348440 & 2 & 9.881 & 2.06 & 0.18 & 0.22 & 8177 & 224 & 1.12 & 181.58 & 18.03 & 38.734 & 1.03 & 44.83 & 43.35 & 44.60 & 43.02 & -- \\
335390527 & -- & 2 & 10.026 & 2.26 & 0.25 & 0.22 & 7899 & 260 & 1.04 & -- & -- & 51.828 & 1.308 & 48.86 & 50.50 & 50.02 & 50.70 & -- \\
335392263 & -- & 2 & 10.007 & 2.20 & 0.24 & 0.25 & 7873 & 180 & 0.99 & 144.24 & 12.15 & 19.763 & 1.074 & 26.75 & 23.37 & 24.90 & 21.83 & 6.88 \\
336321774 & -- & 2 & 10.238 & 2.40 & 0.28 & 0.16 & 7769 & 129 & 0.97 & 221.82 & 55.03 & 19.209 & 1.533 & 28.68 & 21.96 & 23.61 & 20.51 & 6.54 \\
336435989 & -- & 2 & 10.195 & 2.44 & 0.33 & 0.25 & 14129 & 500 & 0.81 & 240.39 & 34.64 & 19.226 & 3.045 & 29.53 & 25.63 & 25.87 & 23.46 & -- \\
336555676 & 336771 & 10 & 10.311 & 2.50 & 0.32 & 0.16 & 7646 & 166 & 1.02 & 53.56 & 5.28 & 32.374 & 0.974 & 28.39 & 30.96 & 29.88 & 31.52 & -- \\
336900245 & -- & 2 & 9.679 & 2.19 & 0.21 & 0.16 & 8040 & 189 & 0.93 & 128.59 & 12.33 & 52.152 & 1.258 & 46.82 & 48.07 & 47.76 & 48.81 & 6.92 \\
339458367 & 230089 & 2 & 10.336 & 2.23 & 0.17 & 0.04 & 8111 & 207 & 0.94 & 70.47 & 9.35 & 57.626 & 1.261 & 54.50 & 55.06 & 54.66 & 55.11 & -- \\
342795098 & 348730 & 2 & 10.213 & 2.18 & 0.14 & -0.01 & 8154 & 152 & 1.15 & 127.80 & 67.69 & 70.626 & 2.48 & 57.82 & 62.94 & 62.19 & 63.48 & 7.35 \\
350993264 & -- & 10 & 10.500 & 2.67 & 0.37 & 0.12 & 7383 & 127 & 0.99 & 17.11 & 3.31 & 27.434 & 0.509 & 27.88 & 27.73 & 29.83 & 28.30 & -- \\
350993729 & -- & 2 & 10.003 & 2.17 & 0.17 & 0.10 & 8375 & 139 & 1.05 & 96.87 & 8.91 & 51.321 & 0.758 & 55.46 & 54.11 & 56.07 & 54.16 & -- \\
352553182 & 239407 & 2 & 9.517 & 2.06 & 0.18 & 0.23 & 8795 & 440 & 1.21 & -- & -- & 46.529 & 0.723 & 42.11 & 43.12 & 42.76 & 43.37 & 6.59 \\
353100191 & -- & 2 & 10.280 & 2.33 & 0.26 & 0.17 & 7814 & 181 & 0.98 & -- & -- & 16.981 & 1.456 & 25.49 & 19.11 & 20.50 & 16.74 & 7.53 \\
354062527 & 231034 & 2 & 10.171 & 2.20 & 0.18 & 0.08 & 8058 & 606 & 1.13 & -- & -- & 52.922 & 1.871 & 50.34 & 51.48 & 50.70 & 51.44 & 6.94 \\
358141782 & 336999 & 10 & 10.491 & 2.60 & 0.38 & 0.21 & 7212 & 181 & 0.86 & 119.45 & 40.35 & 53.488 & 0.863 & 50.98 & 51.36 & 50.85 & 50.74 & 6.96 \\
364492329 & -- & 10 & 10.324 & 2.61 & 0.37 & 0.20 & 7736 & 138 & 0.95 & 97.30 & 16.25 & 46.034 & 1.881 & 33.97 & 39.20 & 37.74 & 42.22 & -- \\
365756191 & -- & 2 & 10.009 & 2.19 & 0.19 & 0.13 & 8242 & 123 & 3.35 & 69.34 & 9.26 & 57.227 & 2.205 & 53.53 & 55.66 & 54.32 & 55.71 & 6.79 \\
375327152 & 239551 & 2 & 9.882 & 2.36 & 0.27 & 0.16 & 9478 & 513 & 1.45 & -- & -- & 50.593 & 3.808 & 41.16 & 48.21 & 48.24 & 49.72 & 6.78 \\
384518807 & -- & 2 & 10.402 & 2.49 & 0.30 & 0.11 & 7914 & 476 & 1.14 & -- & -- & 24.701 & 1.004 & 31.00 & 27.90 & 29.32 & 26.40 & 6.64 \\
390059249 & -- & 2 & 10.079 & 2.20 & 0.18 & 0.08 & 8348 & 137 & 1.07 & 152.43 & 11.90 & 52.756 & 1.822 & 51.28 & 51.52 & 51.52 & 51.74 & 7.33 \\
392932308 & -- & 10 & 9.160 & 1.88 & 0.14 & 0.29 & 8581 & 167 & 0.91 & 260.30 & 16.96 & 25.119 & 0.333 & 32.00 & 30.07 & 27.83 & 27.43 & -- \\
403174967 & 337913A & 2 & 9.983 & 2.13 & 0.20 & 0.19 & 8595 & 483 & 0.99 & 115.44 & 22.17 & 60.121 & 0.475 & 53.63 & 55.22 & 54.32 & 55.52 & 6.84 \\
403533149 & -- & 2 & 10.532 & 2.41 & 0.28 & 0.16 & 7959 & 162 & 1.00 & 199.75 & 27.12 & 21.951 & 2.172 & 31.36 & 30.37 & 32.24 & 30.62 & 6.47 \\
416930588 & -- & 2 & 10.383 & 2.56 & 0.33 & 0.13 & 7457 & 128 & 0.92 & 69.38 & 9.94 & 20.085 & 0.489 & 25.54 & 24.78 & 25.69 & 24.89 & -- \\
424629249 & 336324 & 2 & 10.314 & 2.23 & 0.26 & 0.27 & 7960 & 133 & 1.14 & 212.71 & 42.30 & 17.658 & 1.242 & 28.91 & 26.96 & 27.33 & 25.25 & 6.54 \\
429216438 & 239518 & 2 & 9.769 & 2.30 & 0.19 & 0.02 & 8423 & 661 & 0.99 & -- & -- & 47.095 & 1.119 & 46.78 & 48.11 & 47.54 & 48.31 & 7.07 \\
\\
\bottomrule
\end{tabular}
\end{table}
\end{landscape}

\pagebreak

\begin{table*}
\caption{Excerpt from the table of pulsation frequencies (full table available online), showing the first and last 3 rows. The frequencies $f_i$ and amplitudes $a_i$ of the five strongest peaks above 10\,d$^{-1}$ are shown, sorted by amplitude (descending). Where fewer than five peaks were significant, the remainder are empty.}
\label{tab:modes}
\begin{tabular}{rcccccccccc}
\toprule
\multicolumn{1}{c}{TIC} & $f_1$ & $f_2$ & $f_3$ & $f_4$ & $f_5$ & $A_1$ & $A_2$ & $A_3$ & $A_4$ & $A_5$ \\
 & d$^{-1}$ & d$^{-1}$ & d$^{-1}$ & d$^{-1}$ & d$^{-1}$ & ppt & ppt & ppt & ppt & ppt \\
\midrule
$ 7992735 $&$ 19.264 $&$ 20.236 $&$ 18.742 $&$ 19.632 $&$ 17.773 $&$ 1.003 $&$ 0.891 $&$ 0.46 $&$ 0.386 $&$ 0.376 $\\
$ 8598196 $&$ 21.115 $&$ 20.823 $&$ 19.831 $&$ 22.318 $&$ 21.532 $&$ 0.505 $&$ 0.426 $&$ 0.308 $&$ 0.27 $&$ 0.261 $\\
$ 10339309 $&$ 20.773 $&$ 19.714 $&$ 39.25 $&$ 24.753 $&$ 34.409 $&$ 1.326 $&$ 0.622 $&$ 0.402 $&$ 0.273 $&$ 0.269 $\\
$ 469421586 $&$ 37.122 $&$ 41.574 $&$ 53.862 $&$ - $&$ - $&$ 0.049 $&$ 0.013 $&$ 0.009 $&$ - $&$ - $\\
$ 522220718 $&$ 51.021 $&$ 41.092 $&$ 47.852 $&$ 54.772 $&$ 51.421 $&$ 0.252 $&$ 0.177 $&$ 0.117 $&$ 0.09 $&$ 0.075 $\\
$ 1979640514 $&$ 28.119 $&$ 29.605 $&$ 42.674 $&$ 34.111 $&$ 22.019 $&$ 0.596 $&$ 0.436 $&$ 0.269 $&$ 0.265 $&$ 0.181 $\\
\bottomrule
\end{tabular}
\end{table*}

\bsp	
\label{lastpage}

\end{document}